\documentclass[journal]{IEEEtran}

\usepackage{xcolor}
\usepackage{float}
\usepackage{graphicx}
\usepackage{amsmath}
\usepackage{color}
\usepackage{epsfig}
\usepackage{amsfonts}
\usepackage{amssymb}
\usepackage{amsthm}
\usepackage{enumerate}
\usepackage{comment}
\usepackage{caption}
\usepackage{subcaption}

\begin{document}
\title{Deep Channel Prediction: A DNN Framework for Receiver Design in Time-Varying Fading 
Channels}
\author{Sandesh Rao Mattu, Lakshmi Narasimhan T, and A. Chockalingam \\
\thanks{
© 2022 IEEE. Personal use of this material is permitted. Permission from IEEE must be obtained for all other uses, in any current or future media, including reprinting/republishing this material for advertising or promotional purposes, creating new collective works, for resale or redistribution to servers or lists, or reuse of any copyrighted component of this work in other works.
}}
\maketitle

\begin{abstract}
In time-varying fading channels, channel coefficients are estimated using pilot symbols that are transmitted every coherence interval. For channels with high Doppler spread, the rapid channel variations over time will require considerable bandwidth for pilot transmission, leading to poor throughput. In this paper, we propose a novel receiver architecture using deep recurrent neural networks (RNNs) that learns the channel variations and thereby reduces the number of pilot symbols required for channel estimation. Specifically, we design and train an RNN to learn the correlation in the time-varying channel and predict the channel coefficients into the future with good accuracy over a wide range of Dopplers and signal-to-noise ratios (SNR). The proposed training methodology enables accurate channel prediction through the use of techniques such as teacher-force training, early-stop, and reduction of learning rate on plateau. Also, the robustness of prediction for different Dopplers and SNRs is achieved by adapting the number of predictions into the future based on the Doppler and SNR. Numerical results show that good bit error performance is achieved by the proposed receiver in time-varying fading channels. We also propose a data decision driven receiver architecture using RNNs that further reduces the pilot overhead while maintaining good bit error performance.
\end{abstract}

\vspace{-1mm}
\section{Introduction}
\IEEEPARstart{N}{eural} networks have found applications in a wide range of fields. They are being increasingly used for inference tasks like regression and classification. With the advent of libraries available for training, it has become easier than ever to train and deploy application specific neural networks. This is aided by the availability of hardware tailor made for training neural networks. The time required to train neural networks, even complex ones having large number of parameters, has reduced drastically. The networks also have the advantage that once trained, they are computationally efficient when compared to the conventional optimal algorithms. In the field of communications, neural networks have been employed in a wide range of problems in the physical layer design \cite{ref1},\cite{ref2},\cite{ref2x}. Some of them include design of codes using neural networks \cite{ref3}, decoding algorithms via deep learning \cite{ref2a},\cite{ref2b}, signal detection \cite{ref3}-\cite{ref3cx}, channel estimation \cite{ref4}-\cite{ref4b}, beamforming and precoding \cite{ref4bx},\cite{ref4by}, and autoencoder based transceiver designs for fading channels \cite{sandesh}. In addition to classification and regression problems, neural networks have been used in prediction tasks like clinical prediction \cite{suresh} and caption prediction for images \cite{bengio}. 

A key problem in wireless communications is channel estimation.  Specifically, the receiver needs an estimate of the channel fade coefficient in a given coherence interval of the channel for reliable decoding of  data symbols in that coherence interval. Towards this, known symbols called pilots are sent over the channel to the receiver. The receiver estimates the channel coefficient from the received pilots and uses it for decoding the data symbols. The transmitter sends periodic pilot symbols for channel estimation in every coherence interval. The pilot symbols take up considerable portion of the available bandwidth if the coherence time of the channel is small. 

In this paper, we consider a scenario where there is mobility at the receiver and/or at the transmitter. The relative motion between the transmitter and receiver introduces Doppler spread in the channel and this leads to the channel fade coefficients being time correlated. We aim to use neural networks to take advantage of this correlation among the fade coefficients to reduce the pilot resources for communication. We achieve this by training neural networks that can learn the temporal dependency in the fading process and use this knowledge to predict future values of the fade coefficients. This prediction of future fade coefficients allows pilot symbols to be sent less often, leading to increased data throughput. Our new contribution in this paper is that we propose a novel receiver architecture that uses recurrent neural networks (RNN) that perform deep channel prediction and signal detection in time-varying fading channels. To our knowledge, an RNN-based channel prediction approach for the design of robust receivers in time-varying fading channels has not been reported. The new contributions in this paper can be summarized as follows. 
\begin{itemize}
\item First, we design and train an RNN to learn the correlation in the time-varying fading channel and predict the channel coefficients into the future with good accuracy over a wide range of Dopplers and signal-to-noise ratios (SNR). The proposed training methodology enables accurate channel prediction through the use of techniques such as teacher-forced training, early-stop, and reduction of learning rate on plateau. The robustness of prediction for different Dopplers and SNRs is achieved by adapting the number of predictions into the future based on the Doppler and SNR. Our numerical results show that good bit error rate (BER) performance is achieved by the proposed receiver in time-varying fading channels. 
\item Next, we propose a data decision driven receiver architecture using RNNs that further reduces the pilot overhead while maintaining good bit error performance. 
\end{itemize}
The achieved robustness in the receiver performance over a range of Doppler and SNR conditions illustrates that the proposed RNN-based channel prediction approach is a promising approach for receiver design in time-varying fading channels.

The rest of the paper is organized as follows. In Sec. \ref{sec2}, we present the considered system model and a brief background on deep neural network architectures used in this paper. In Sec. \ref{sec3}, we present the proposed deep channel predictor, its architecture, training methodology, and performance. The proposed adaptive channel prediction scheme and its performance are also presented in this section. In Sec. \ref{sec4}, we present the proposed data decision driven architecture and its performance. Conclusions are presented in Sec. \ref{sec5}.

\section{System Model}
\label{sec2}
Consider a point-to-point wireless communication system with a single antenna transmitter and receiver. The channel between the transmitter and receiver is a time-varying fading channel. The information symbols are chosen from an $M$-ary constellation. Let $x(t)$ be the transmit signal at the $t$th time instant. The channel fade coefficient at the $t$th time instant is denoted by $h(t)$. Let $y(t)$ be the received signal at the receiver and $n(t)$ be the additive noise. Now, $y(t)$ can be written as
\begin{align}
y(t) = h(t)x(t) + n(t).
\label{channel_model}
\end{align}
The channel fade coefficients are statistically modelled by a circularly symmetric complex Gaussian random variable with mean 0 and variance 1, i.e., $h\sim\mathcal{CN}(0, 1)$.
The additive white Gaussian noise (AWGN), $n(t)$, is modelled as $n\sim\mathcal{CN}(0, \sigma^2)$, where $\sigma^2$ is the variance of the noise.

We consider a mobile communication scenario where there is relative motion between the transmitter and receiver. This introduces Doppler spread in the channel due to time selectivity and the channel fades $h(t)$ become temporally correlated. The correlation in $h(t)$ depends on several factors such as scatterers in the propagation environment, relative velocity between the transmitter and receiver, etc. The power spectral density (PSD) of $h(t)$ is non-zero in the interval $[-f_D^{\text{max}}, f_D^{\text{max}}]$, where $f_D^{\text{max}}$ is the maximum Doppler frequency given by \cite{clarke},\cite{smith}
\begin{equation}
f_D^{\text{max}} = \frac{f_c v}{c}.
\label{eq_fd}
\end{equation}
In \eqref{eq_fd},
$v$ is the maximum relative velocity between the transmitter and the receiver, $f_c$ is the carrier frequency and $c$ is the speed of light. Therefore, the Doppler spread of the channel is given by $2f_D^{\max}$. For a low Doppler spread, the channel changes slowly over time, while a high value of Doppler spread indicates that the channel varies rapidly with time. The coherence time ($T_c$) of the channel is inversely proportional to the Doppler spread, $T_c \propto 1/f_D^{\max}$. In order to detect the transmitted signal $x(t)$ from $y(t)$, the value of $h(t)$ has to be estimated at the receiver. In each  transmission block spanning one coherence time, the channel gain is estimated and employed for detection of the data signal transmitted in that coherence block.

Typical wireless communication systems transmit one or more pilot symbols in each coherence block to estimate the channel coefficients. Let $T_p$ and $T_d$ be the duration of pilot transmission and data transmission, respectively, in a coherence block, i.e., $T_c=T_p+T_d$. Let $p(t)$ be the pilot signal transmitted at the $t$th time instant. The signal received during the pilot transmission phase can be written as
\begin{equation}
y_p(t)=h(t)p(t) + n(t).
\label{fading_model}
\end{equation}
The linear minimum mean square error (LMMSE) estimate of the channel coefficient that achieves the Cramer-Rao lower bound is given by \cite{van2004optimum},\cite{tse} 
\begin{align}
 \widehat{h}(t)=\frac{y_p(t)\vert p(t)\vert^2}{p(t)\left(\vert p(t) \vert^2 + \sigma^2 \right)}.
\label{lmmse_est}
\end{align}

The transmission of pilots reduces the spectral efficiency and throughput of the communication system. That is, $\frac{T_p}{T_c}$ fraction of the channel-uses do not carry data. The efficiency of channel usage is defined as \begin{equation}
\eta=1-\frac{T_p}{T_c}.
\end{equation}
For a fixed number of pilots per coherence block, as the coherence time decreases, $\eta$ also decreases. High mobility wireless communication channels may require large amount of bandwidth to be used for pilot transmission, which, in turn, adversely reduces the achievable data rate and system capacity.

Since the time-varying channel coefficients are temporally correlated, the number of pilots transmitted to estimate the channel coefficients can be reduced by learning this correlation model and using the learnt model in channel estimation. As the correlation model could be different for different channel geometries, a statistical solution to this problem may not be robust. Therefore, we propose to employ a deep learning based solution to learn the channel correlation model and predict the channel coefficients into the future to reduce the pilot transmissions. Towards this, we employ recurrent neural networks (RNN) and fully connected neural networks (FCNN) to construct the proposed deep channel predictor and the receiver. Consequently, in the following subsection, we present a brief background on deep neural networks, focusing on FCNNs and RNNs. 

\subsection{Deep neural networks}
\label{sec2a}
The deep neural network architectures that we employ are FCNNs and RNNs.

\subsubsection{Fully connected neural networks}
\begin{figure}[ht]
\centering
\includegraphics[height=7cm,width=7cm]{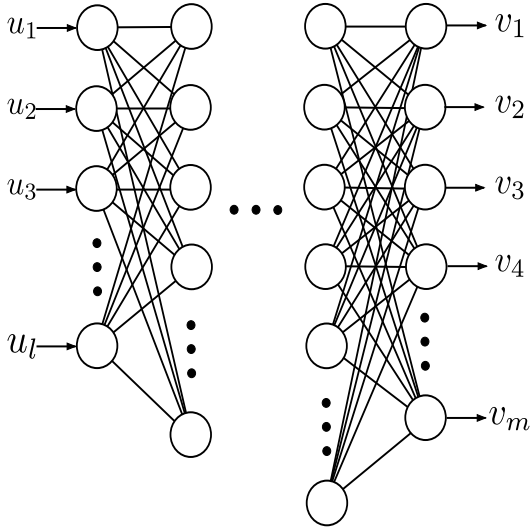}
\caption{Architecture of a fully connected neural network.}
\label{fig:fcnn}
\end{figure}

A deep FCNN consists of multiple layers of neurons. Every neuron in a layer is connected to all neurons in the adjacent layers, thus forming a fully connected network. The architecture of a deep FCNN is illustrated in Fig. \ref{fig:fcnn}. Deep FCNNs have been known to be suitable for learning or approximating any linear/non-linear function for tasks such as detection or estimation \cite{jordan}. For data that is temporally correlated, RNNs are known to provide better learning performance than FCNN \cite{elman}.

\subsubsection{Recurrent neural networks}
\begin{figure}[ht]
\centering
\includegraphics[width=0.9\linewidth]{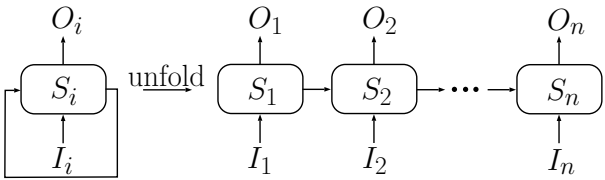}
\caption{Architecture of a recurrent neural network.}
\label{fig:rnn}
\end{figure}

A deep RNN can be constructed by the repetition of a one or more blocks over time, where a single block consists of multiple trainable parameters. That is, the output of a single block is fed back recursively, thus enabling the network to have memory and learn temporal correlation in the input data. This is referred to as `time unfolding' of the network. The architecture of a deep RNN is illustrated in Fig. \ref{fig:rnn}.

In Fig. \ref{fig:rnn}, the left portion shows the recurring unit of the RNN and the right portion illustrates the unfolding over time. The input and output of the RNN at the $i$th time instant are represented by $I_i$'s and $O_i$'s, respectively. $S_i$'s are referred to as hidden states of the RNN. There are three trainable weight matrices and two trainable biases in such an RNN. A weight matrix $\mathbf{W}_{SI}$ is employed in the link between $I_i$ and $S_i$. Similarly, weight matrices $\mathbf{W}_{OS}$ and $\mathbf{W}_{SS}$ are employed in the links between $S_i$ and $O_i$, and between $S_i$ and $S_{i-1}$, respectively. The biases $\mathbf{b}_S$ and $\mathbf{b}_O$ are added when computing $S_i$ and $O_i$, respectively. The matrices and biases are the same across all unfolding, i.e., the entries of the matrices and biases are not a function of the time.

There are several implementations of RNN. In this paper, we make use of an implementation known as the long short-term memory (LSTM) \cite{lstm}.
\begin{figure}
\centering
\includegraphics[width=0.8\linewidth]{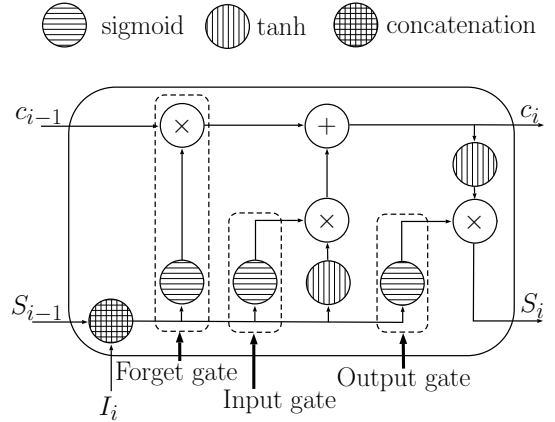}
\caption{Recurrent unit of the LSTM architecture.}
\label{fig:lstm_arch}
\end{figure}
The block diagram of the recurrent unit of the LSTM architecture of RNN is shown in Fig. \ref{fig:lstm_arch}. This architecture consists of three gates. These gates learn the temporal information that are relevant and pass it to the next iteration. In each gate, a sigmoid function is applied that restricts the output to values between $0$ and $1$. The output of the activation are then multiplied to decide which part of the information is relevant. During training, the weights are updated such that the relevant information gets a larger weight which yields a value close to $1$ after the sigmoid function.
In Fig. \ref{fig:lstm_arch}, the variable $c_{i}$, called the cell state, is made available to all unfolded blocks. The variable $S_i$ refers to the hidden state of the cell and $I_i$ is the input to the cell. In our setup, the input $I_i$ corresponds to either the channel estimates from \eqref{lmmse_est} or the fed back prediction values (see Fig. \ref{fig:predictor_block}). The $c_i$'s and $S_i$'s are updated at each stage $i$ using $I_i$'s. However, the information that is passed on to the next iteration depends on the gate values. We use LSTM implementation of the RNN because, as opposed to the basic RNN implementation, LSTMs are able to learn correlation model in long time varying sequences \cite{lstm}.

We use PyTorch machine learning library for the implementation, training, and testing of all the neural networks proposed in this paper \cite{pytorch}. We use the Nvidia Titan RTX GPU platform to carry out all the simulations.

\section{Proposed Deep Channel Predictor}
\label{sec3}
In this section, we present the proposed RNN based deep channel predictor, its architecture, training methodology, and performance. The channel predictor uses the received pilot symbols to learn the channel variation model and predict the future channel coefficients. 

\subsection{Architecture}
\label{sec3a}
\begin{figure}
\centering
\includegraphics[width=0.9\linewidth]{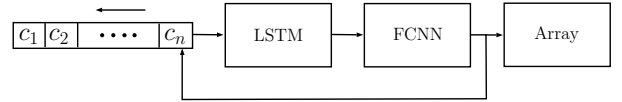}
\caption{Block diagram of the channel predictor neural network.}
\label{fig:predictor_block}
\end{figure}

The proposed deep channel predictor consists of two prediction networks, one each for predicting the real and imaginary parts of the channel coefficients. The architecture for these networks are the same and they are trained separately. The deep channel predictor network consists of an LSTM network and an FCNN. The block diagram of the proposed deep channel predictor is shown in Fig. \ref{fig:predictor_block}.

\begin{figure}
    \centering
    \includegraphics[width=\linewidth]{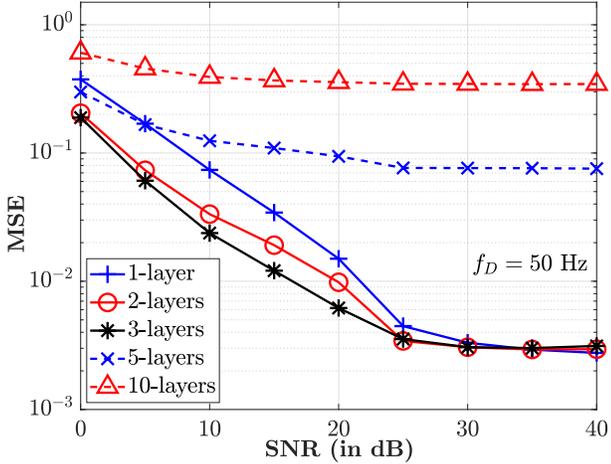}
    \caption{Mean square error of predictions made by predictor network for different number of layers in the LSTM network.}
    \label{fig:layers_vs_mse}
\end{figure}

\begin{figure}
    \begin{subfigure}{0.49\linewidth}
        \centering
        \includegraphics[width=4.75cm, height=4.5cm]{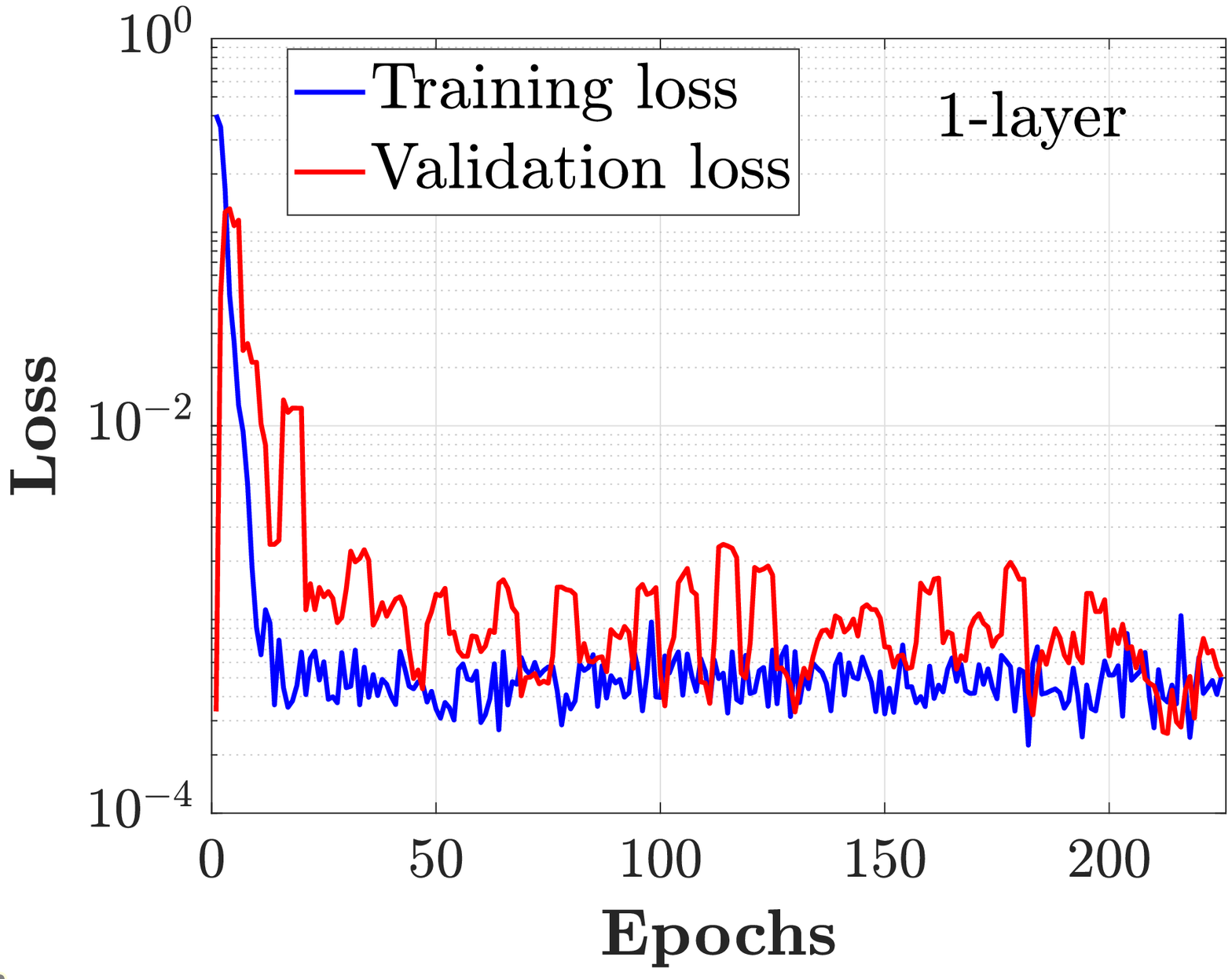}
        \caption{1-layer}
        \label{fig:loss_1_layer}
    \end{subfigure}
    \begin{subfigure}{0.49\linewidth}
        \centering
        \includegraphics[width=4.75cm, height=4.5cm]{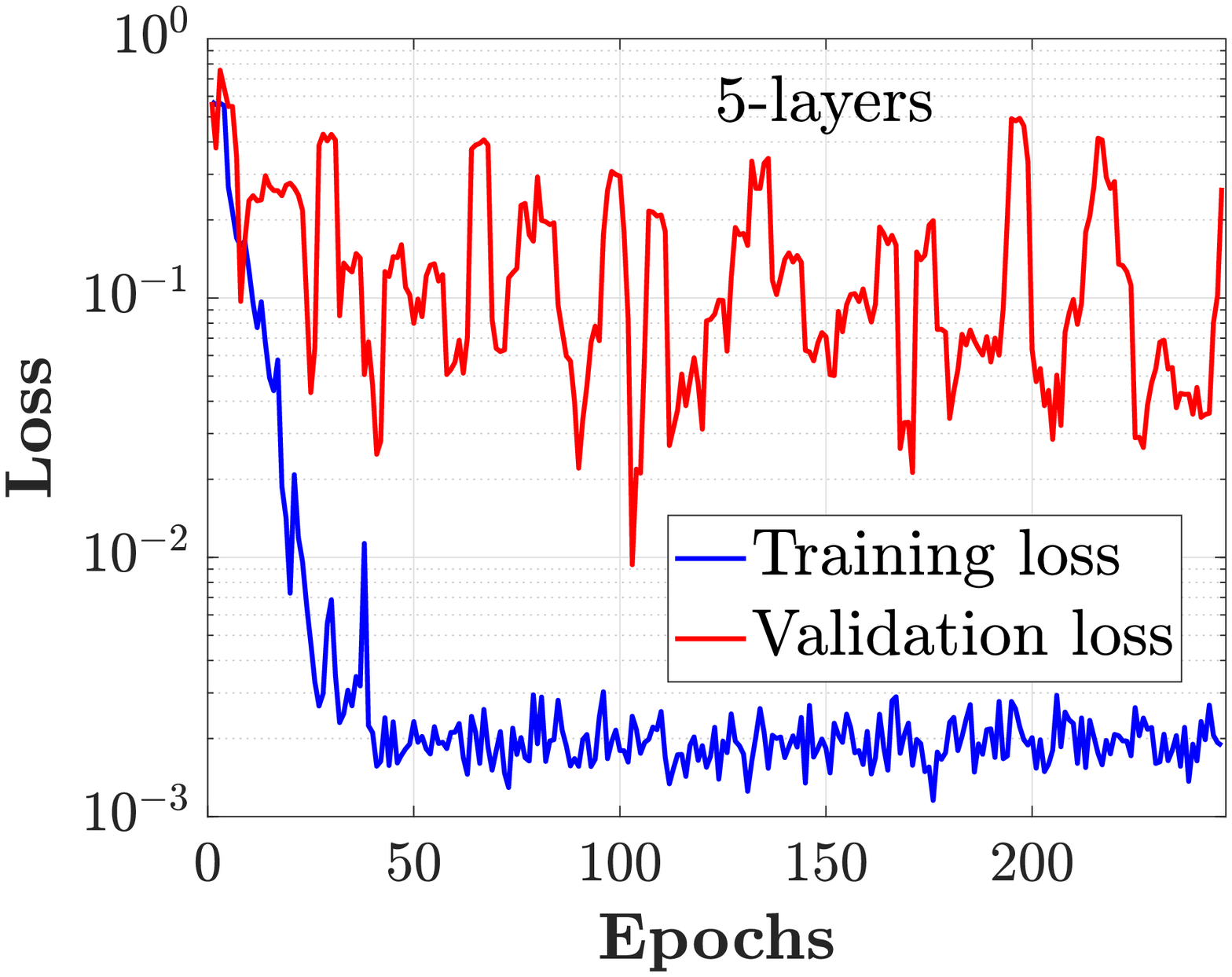}
        \caption{5-layers}
        \label{fig:loss_5_layers}
    \end{subfigure}
    \caption{Training and validation loss trajectory for 1-layer and 5-layers LSTM architectures.}
    \label{fig:loss_vs_layers}
\vspace{-4mm}
\end{figure}

\begin{table}[t]
\centering
\begin{tabular}{|l|l|}
\hline
\bf{Parameter} & \bf{Value} \\
\hline
Number of layers & 1 \\
\hline
Input dimension & 1 \\
\hline
Hidden units & 100 \\
\hline
Output dimension ($l$) & 100  \\
\hline
Direction & Uni-directional \\
\hline
\end{tabular}
\vspace{2mm}
\caption{Parameters of LSTM layer of channel predictor.}
\label{tab:hyper_paramters_LSTM}
\vspace{-4mm}
\end{table}

The purpose of using LSTM is two fold. First, the LSTM is capable of identifying temporal correlations in the inputs and learning a correlation model. Second, the LSTM can leverage the learnt correlation model to make predictions that obey the model. Also, we choose a single layer LSTM architecture for the predictor network based on the following performance evaluation.   Figure \ref{fig:layers_vs_mse} plots the mean square error (MSE) performance of the trained predictor network as a function of SNR, for different number of layers in the LSTM architecture. It is seen that the MSE of the predictions improves in the low to  mid SNR regime when the number of layers is  increased from 1 to 3. However, with further increase in the number of layers, the MSE performance of the network degrades, which  can  be  attributed  to  the phenomenon of over-fitting, wherein a network learns to perform well only on the data set used in training \cite{lawrence}. This is illustrated in Figs. \ref{fig:loss_1_layer} and \ref{fig:loss_5_layers} which show the training/validation loss performance for 1-layer LSTM and 5-layer LSTM, respectively. It is seen that both 1-layer and 5-layer LSTMs show convergence to small training loss values (indicating successful training). However, in the validation phase (where data not in the training data set is used for validation) the validation loss in 5-layer LSTM does not show convergence to small values (indicating over-fitting), while 1-layer LSTM achieves convergence to small loss values in validation phase as well. Although LSTM  architecture with 3-layers has the best MSE performance, the improvement it  offers over 1-layer LSTM architecture is not significant compared to the complexity it introduces. For example, the number of parameters in the predictor network with 1-layer, 2-layer, and 3-layer LSTMs are 41301, 122101, and 202901, respectively. The five-fold increase in the number of parameters when compared to the 1-layer LSTM makes the 3-layer architecture to be slower and  harder to train than the 1-layer counterpart. Further, above 25 dB, the MSE performance of the predictor network with 1-layer, 2-layers, and 3-layers LSTM architectures are almost similar. Therefore, we choose 1-layer LSTM architecture with the parameters listed in Table \ref{tab:hyper_paramters_LSTM} for the predictor network throughout this paper for its simplicity and reasonably good MSE performance.

The FCNN layer is employed to reduce the output of the LSTM layer to the required dimension. In our setup, the data from the output of the LSTM has a dimension of 100, which is to be reduced to a dimension 1 indicating a single channel prediction. However, picking one dimension arbitrarily may not yield the best solution. The FCNN takes the 100-dimensional data as input, and during training assigns large weights to those outputs which have a potentially higher bearing on the prediction value as compared to the rest. This can improve the performance of the predictions made by the setup. The FCNN layer parameters are listed in Table \ref{tab:hyper_paramters_FCNN_pred}.

Figure \ref{fig:predictor_block} shows the block diagram that depicts the working of the channel predictor. The predictor network expects an $n$-length sequence of channel coefficients as input. The working of the network is divided into two phases, namely, the initial estimation phase and the subsequent prediction phase. In the estimation phase, $n$ pilots are transmitted in $n$ coherence intervals. The LMMSE channel estimates from these transmitted pilots are obtained using \eqref{lmmse_est}. These estimates are used to initialize the entries of the input vector $\mathbf{c}=[c_1 \ c_2 \ \cdots \ c_n]$ with entries arranged chronologically, $c_1$ being the least recent estimate and $c_{n}$ being the most recent. This initialized vector is provided as the input to the LSTM network. The entries of $\mathbf{c}$ reflect the correlation among the channel coefficients, and are used by the LSTM network to predict the channel coefficients in the subsequent coherence intervals. The output of the LSTM network is fed to the FCNN layer.

\begin{table}
\vspace{6.0mm}
\centering
\begin{tabular}{|l|l|}
\hline
\bf{Parameter} & \bf{Value} \\
\hline
Input dimension ($l$) & 100 \\
\hline
output dimension ($m$) & 1 \\
\hline
Activation function & Linear \\
\hline
Number of layers & 1  \\
\hline
\end{tabular}
\vspace{2mm}
\caption{Parameters of FCNN layer of channel predictor.}
\label{tab:hyper_paramters_FCNN_pred}
\vspace{-4mm}
\end{table}

\begin{figure}[t]
\centering
\includegraphics[width=0.8\linewidth]{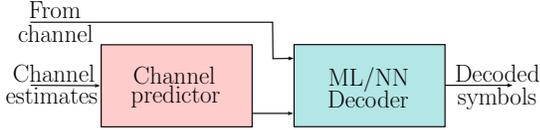}
\caption{Block diagram of the channel predictor aided receiver.}
\label{fig:without_feedback_block_dia}
\end{figure}

The FCNN layer produces one channel prediction at its output. This output is the prediction for one-step into the future, i.e., this output is the predicted coefficient for the coherence interval next to the coherence interval for which the most recent estimate $c_{n}$ was obtained. This concludes the initial estimation phase. Subsequently, in the prediction phase, the input vector $\mathbf{c}$ is left shifted so that $c_1$ is flushed out and the previously obtained prediction value is used to fill the vacant $c_{n}$ space after the left shift operation. The input of the LSTM is thus updated with the most recent prediction. A procedure similar to that in the estimation phase is followed again to obtain the channel prediction corresponding to the next coherence interval. This process is repeated for as many times as the number of predictions required. The predictions thus made by the network are stored in an array. At the end of required number of predictions, the array is used to decode the transmitted symbols. The architecture is therefore flexible in the sense that it allows for dynamic adjustment of the number of channel predictions.

The block diagram of the overall channel predictor aided receiver is shown in Fig. \ref{fig:without_feedback_block_dia}. The channel predictor block is followed by a data decoder. The data decoder can be maximum likelihood (ML) decoder or a neural network (NN) based decoder. We will use ML decoding for transmission schemes that decode symbol by symbol, due to low ML decoding complexity in such cases. For block transmission schemes which require joint decoding of symbols (e.g., cyclic prefix single carrier (CPSC) scheme in Sec. \ref{sec3g}), we will use NN-based decoder approach.

Further, the channel predictor and the NN decoder can be trained together as a single network, as it would alleviate the need for a separate decoder. However, we keep the training for the predictor and the decoder separate with the intention of having a universal predictor network. That is, once trained, the predictor network can be used in conjunction with any decoder. On the other hand, if training is done for the predictor and decoder together, there is a need to train and store multiple models, each corresponding to a different decoder.

\begin{figure*}
\centering
\begin{subfigure}{0.5\linewidth}
\centering
\includegraphics[width=\linewidth]{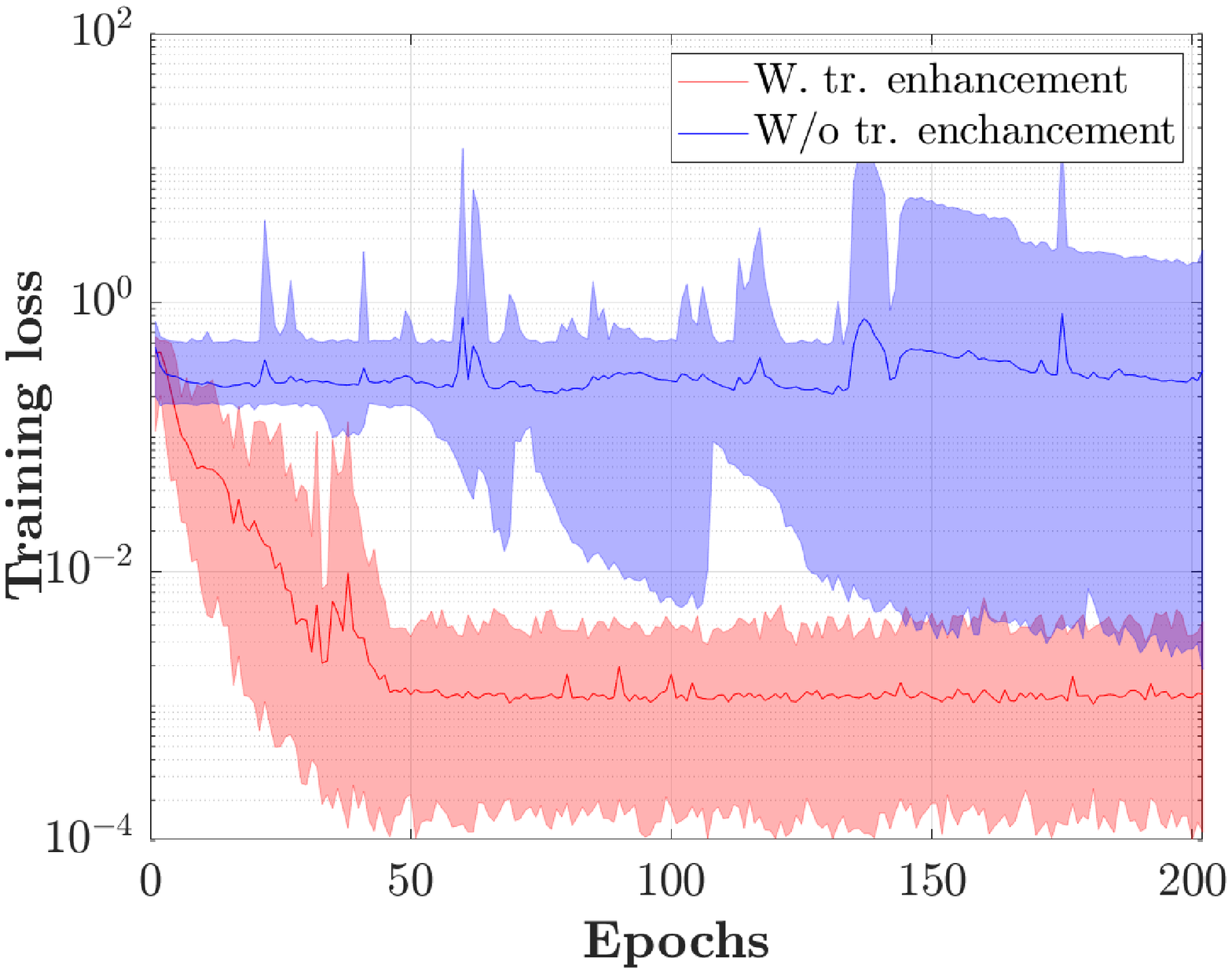}
\caption{Training loss}
\label{fig:train_loss}
\end{subfigure}%
\begin{subfigure}{0.5\linewidth}
\centering
\includegraphics[width=\linewidth]{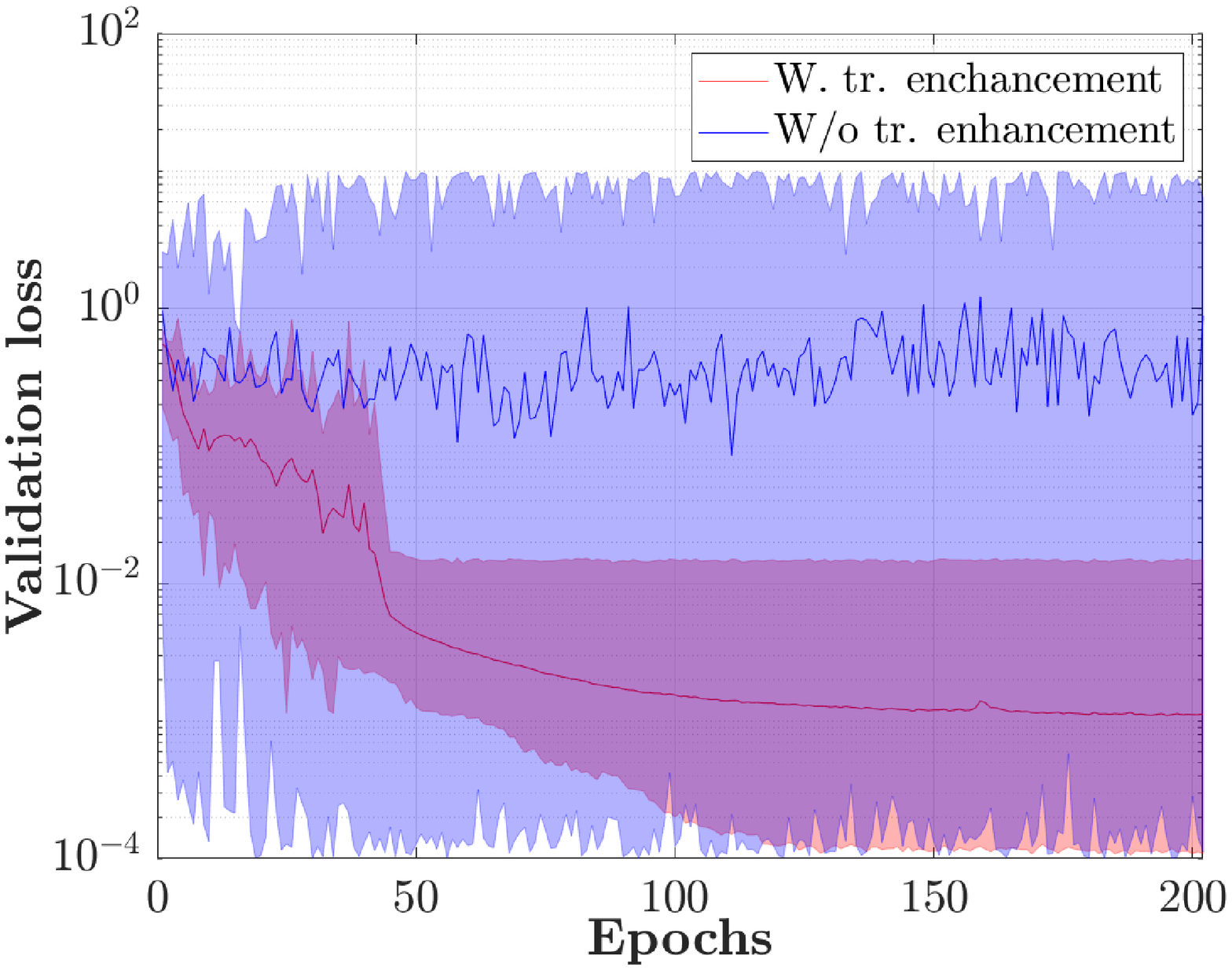}
\caption{Validation loss}
\label{fig:val_loss}
\end{subfigure}%
\caption{Comparison of training and validation loss trajectories as a function of epochs with and without the training enhancement features.}
\label{fig:loss_vs_epoch}
\vspace{-2mm}
\end{figure*} 

\subsection{Training methodology}
\label{sec2b}
In this subsection, we describe the training of the channel predictor network. To train the channel predictor network, correlated channel coefficients that mimic a channel with time selectivity are used. This data is generated using the Clarke and Gan's model \cite{clarke}. For a time-selective channel, the auto-correlation function of the fading process $h(t)$ is given by 
\begin{align}
\psi(\Delta t) = J_0(2\pi f_D^{\text{max}}\Delta t),
\label{acf}
\end{align}
where $J_0(.)$ is the modified Bessel function of the first kind and zeroth order. The PSD of $h(t)$, which is also referred to as the Jakes' spectrum, is given by 
\begin{align}
S_H(f) = \frac{\text{rect}\left(\frac{f}{2f_D^{\text{max}}}\right)}{\pi f_D^{\text{max}}\sqrt{1 - \left(\frac{f}{f_D^{\text{max}}}\right)^2}},
\label{PSD}
\end{align}
where rect$(\cdot)$ is the rectangular function defined as
\begin{equation}
\text{rect}(x) = 
\begin{cases}
1, \hfill \ \ x \in [-\frac{1}{2}, \frac{1}{2}] \\
0, \hfill \ \ \text{otherwise.}
\end{cases}
\end{equation}
Specifically, we use the implementation of Clarke and Gan's model given by Smith in \cite{smith} to generate the required data. In our simulations, a coherence block consists of 42 symbols for $f_D=50$ Hz and 11 symbols for $f_D = 100$ Hz. 

To begin the training, the LSTM and FCNN networks in Fig. \ref{fig:predictor_block} are initialized with random or untrained weights. A large number of correlated channel coefficient samples are generated using the implementation mentioned above, so that the network is able to generalize well. The channel coefficients are separated into real and imaginary parts. The real (imaginary) part is used as training data for the predictor network which is set to predict real (imaginary) part of the channel coefficients. While training, the number of predictions made by the network is fixed at 100. The network is trained with 10-length sequence of input data and 100-length sequence of ground truth predictions or expected predictions. Both are obtained from the correlated channel coefficients obtained through Smith's method. Therefore, we sample blocks of length 110 from the generated samples of correlated channel coefficients. The sampled sequence is structured such that the first entry in the sequence is the least recent and the last entry is the most recent. The first 10 entries (from the least recent end) of the sequence are provided as input to the predictor network. The network produces 100-length predictions. These predictions are compared with the last 100 entries of the 110-length sequence by computing a mean square error (MSE) loss function, given by
\begin{align}
L(\mathbf{x}, \hat{\mathbf{x}}, \mathbf{\Theta}) = \mathbb{E}[\mathbf{x} - \hat{\mathbf{x}}(\mathbf{\Theta})]^2.
\label{mse_loss}
\end{align}
In the above equation, $\mathbb{E}[\cdot]$ is the expectation operator, $\textbf{x}$ is the 100-length expected output, and $\hat{\textbf{x}}(\mathbf{\Theta})$ is the 100-length output of the neural network, which is a function of the network parameters $\mathbf{\Theta}$. At each training iteration, the value of \eqref{mse_loss} is computed and the weights are updated so as to minimize the loss function through back propagation. This procedure is repeated for real and imaginary predictions in each iteration. The values of the hyper parameters used for training the channel predictor are given in Table \ref{tab:hyper-parameters training}.

{\it Remark 1:} 
Note that the predictor network is trained at a Doppler frequency of 10 Hz (see Table \ref{tab:hyper-parameters training}). The predictor trained at 10 Hz Doppler is able to predict well over a range of Doppler values (as will be shown in Figs. \ref{fig:mse_vs_nfp} and \ref{fig:mse_vs_dopp} later). This is because the training teaches the LSTM network essentially to observe the underlying correlation model in the input data and leverage the observed model to predict future coefficients. In the 10 Hz case, the network trained at 10 Hz observes that the channel coefficients have strong correlation and it outputs predictions that obey the underlying slow variation model. In the 100 Hz case, the input changes are more abrupt and the same trained network is able to adapt to this underlying fast variation model as well and produce predictions that obey the faster trend.

\begin{table}
\centering
\begin{tabular}{|l|l|}
\hline
\textbf{Hyper parameter} & \textbf{Value} \\
\hline
Starting learning rate & 0.01 \\
\hline
Minimum learning rate & $10^{-8}$ \\
\hline
Epochs &  Min. 200, Max. 1000 \\
\hline
Optimizer & Adam \\
\hline
Loss function & MSE \\
\hline
Batch per epoch & 4500 \\
\hline
Training Doppler frequency & 10 Hz (see {\it Remark 1})\\
\hline
\end{tabular}
\vspace{2mm}
\caption{Hyper parameters used for training channel predictor.}
\label{tab:hyper-parameters training}
\vspace{-4mm}
\end{table}

\subsubsection{Training enhancement features}
\label{sec3b3}
The training method outlined above, by itself, either leads to a large number of iterations before converging (where the loss function assumes a small enough value) or to a condition where the network does not converge at all (where the loss function did not monotonically decrease). This is because in each iteration during the initial part of the training, the prediction made is inaccurate due to untrained weights and the erroneous value is fed back to the input to make another prediction. It is only at the end of 100 predictions that the loss function is evaluated and the back propagation to update weights is performed. Due to the error accumulating at the input, the output might become garbled leading to poor weight updates, resulting in slow convergence or divergence. Therefore, we employ additional techniques while training as enumerated below.
\begin{itemize}
\item {\it Teacher force training:} 
This technique is employed to alleviate the problem mentioned above. Teacher force training involves supplementing the training with the ground truth data. During training, data is fed back from the output of the predictor network to the input. With teacher force training, with small probability, $p$, the ground truth data corresponding to that time instant is supplied from the 100-length expected output. This prevents the input from accumulating error due to inaccurate predictions. In our training setup, we found that a probability of $p=0.2$ works well for quick convergence of the network. This implies that with $1-p=0.8$ probability the prediction made by the network itself is fed back to its input to make further predictions. Since the input is not allowed to deviate uncontrollably from the actual values, this helps the network converge faster.
\item {\it Reduce learning rate on plateau:} 
Learning rate is a hyper parameter that needs to be set while training. The value of the learning rate decides how fast or slow a network learns, by pacing the weight updates. A large learning rate (of the order of $\sim 0.01$) is desirable at the initial stages of training. However, when the loss function hits a plateau a large learning rate may not help the loss function to reduce further. This is because the large value of the learning rate forces large weight updates and may result in unsettling the network from the state it is in. A small learning rate would ensure that the weight updates are small and the would help the network to find the minimum within the plateau. If this does not happen and the loss function continues to maintain the value at plateau, the technique calls for increasing the learning rate back to its original value. In our training setup, we implemented this by reducing learning rate by a factor of 10 every time the loss function value did not reduce for 10 consecutive training iterations. To prevent the learning rate from becoming minuscule, we set the minimum value to be $10^{-8}$. In the process of decreasing the learning rate, if at any stage the value of loss function is found to increase, the learning rate value is reset to its original value.
\item {\it Early stop:}
Yet another problem that is associated with training neural networks is that of over-fitting. Over-fitting is said to occur when the network is allowed to learn for a long time on the available data. This results in a trained model that is tailor made for the training data, but fails to generalize to data beyond those seen while training. That is, the model performs poorly on any data that is not present in the training data. To prevent this from happening we employ a technique called early stop. The early stop technique dictates that the training be stopped when the network is not able to learn any further. This happens when the loss function does not reduce across iterations. We implement this after a minimum of 200 epochs of training. Following this, if the learning rate has already dropped to $10^{-8}$ from the second technique and the loss function does not reduce significantly in the next 50 iterations, we stop training the network. If such a scenario never occurs during training, the training is stopped after 1000 epochs.
\end{itemize}

\subsection{Performance results}
\label{sec3c}
In this subsection, we present simulation results on the training performance, prediction error performance, and BER performance associated with the proposed channel predictor aided receiver developed in the previous subsections. In all the simulations, a fixed 4-QAM symbol is used as a pilot symbol, and the pilot symbol power and the data symbol power are kept the same. In practice, the pilot power is typically kept at the same or a higher level compared to the power in the data symbols.

\subsubsection{Training performance}
\label{sec3c1}

\begin{figure}
\centering
\includegraphics[width=\linewidth]{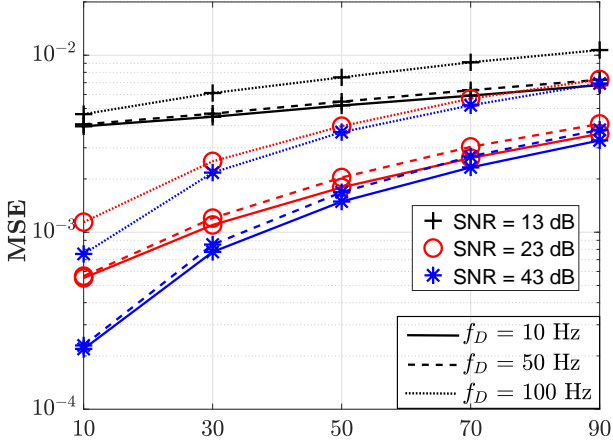}
\caption{Mean square error of predictions as a function of number of predictions for different SNR and $f_D$ values.}
\label{fig:mse_vs_nfp}
\vspace{-2mm}
\end{figure}

\begin{figure*}
\centering
\begin{subfigure}{0.5\linewidth}
\centering
\includegraphics[width=\linewidth]{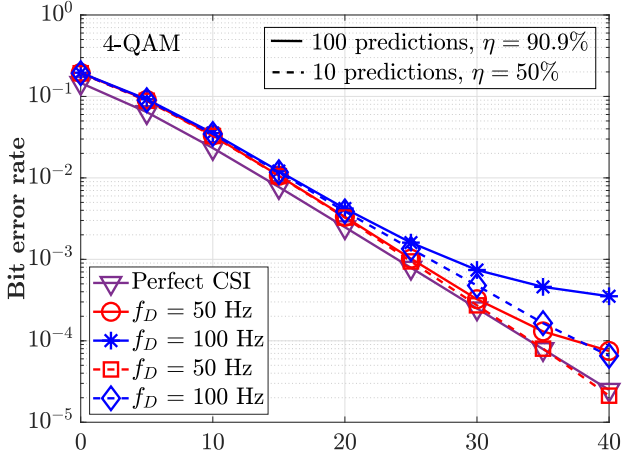}
\caption{4-QAM}
\label{fig:4_qam_fixed}
\end{subfigure}%
\begin{subfigure}{0.5\linewidth}
\centering
\includegraphics[width=\linewidth]{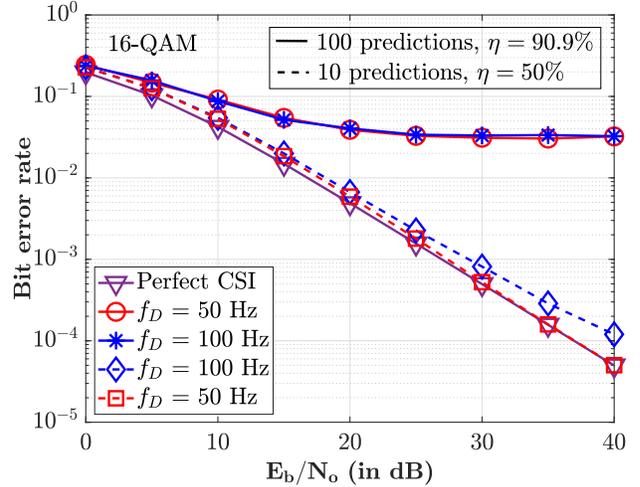}
\caption{16-QAM}
\label{fig:16_qam_fixed}
\end{subfigure}%
\caption{BER Performance of the proposed channel predictor aided receiver with ML decoder for fixed number of predictions ($N$=100, $\eta=90.9\%$ and $N$=10, $\eta=50\%$) at $f_D=50,100$ Hz for 4-QAM and 16-QAM.}
\label{fig:fixed_preds}
\vspace{-2mm}
\end{figure*}

Figure \ref{fig:loss_vs_epoch} shows the training trajectory in terms of training loss and validation loss at each epoch comparing training performed with and without the above mentioned enhancement features. In Figs. \ref{fig:train_loss} and \ref{fig:val_loss}, the plotted line shows the mean of the training loss and validation loss, respectively, while the shaded area around the line is indicative of the variance observed in the losses across training runs. The training loss (MSE loss between the predicted coefficients and the actual coefficients evaluated during training) in the presence of the enhancement features shows convergence at about 50 epochs for the training loss trajectory and about 100 epochs for the validation loss trajectory, after which the loss remains almost constant. This quick convergence is attributed to teacher force training and subsequent consistency in the loss function value is due to the reduced learning rate. Further, as the variance in the validation loss (MSE loss evaluated on data not present in training data) decreases to small values around 200 epochs, the training is stopped and the network parameters are frozen in accordance with the early stop training feature. In contrast, without the enhancement features, the training loss does not seem to converge as it assumes a high value throughout. A similar trend is observed in the validation loss trajectory as well. Without the enhancement features, the network shows large variations in the validation loss and training loss even at 200 epochs, which leads to slow convergence. Figure \ref{fig:loss_vs_epoch}, therefore, demonstrates the effectiveness of the enhancement features in attaining faster convergence.

\subsubsection{Prediction error performance}
\label{sec3c2}
Figure \ref{fig:mse_vs_nfp} shows the MSE performance of predictions as a function of number of future predictions made by the channel predictor. The plots are obtained for $f_D = 10, 50, $ and 100 Hz. The following observations can be made from Fig. \ref{fig:mse_vs_nfp}. First, the MSE performance is found to improve with increasing SNR, which is expected. Next, for a given SNR and $f_D$, increasing the number of future predictions increases the MSE. As the number of predictions is increased at a given SNR and $f_D$, more errors are accumulated which explains the observed trend. Therefore, choosing the right number of future predictions becomes crucial to ensure robustness across different values of Dopplers and SNRs. For a given SNR and number of predictions, the MSE curves for different $f_D$ values are close.

\subsubsection{BER performance}
\label{sec3c3}
In Figs. \ref{fig:4_qam_fixed} and \ref{fig:16_qam_fixed}, we demonstrate the BER performance achieved by the proposed channel prediction aided receiver with ML decoder for 4-QAM and 16-QAM, respectively. Performance with perfect channel state information (CSI) is also plotted for comparison. We consider two scenarios to demonstrate the effect of number of future predictions on the BER performance. The first is a greedy scenario (with respect to bandwidth efficiency), where we set the number of predictions to be fixed at 100 across different $E_b/N_0$ and $f_D$ values. This corresponds to a bandwidth efficiency of 90.9\%. The second is a conservative scenario (with respect to MSE of predictions), where the number of predictions is fixed at 10 instead of 100, corresponding to a bandwidth efficiency of 50\%. The following observations can be made. First, it can be seen that in the conservative scenario with 10 predictions, the achieved BER performance is very close to the ideal performance with perfect CSI for both 4-QAM and 16-QAM with $f_D=50, 100$ Hz. Second, although the greedy scenario achieves good bandwidth efficiency, the BER performance takes a hit. The performance gap between the greedy and the conservative scenarios is more for the channel with a higher Doppler. While the performance hit in the greedy scenario is not very significant in the 4-QAM case, it is quite severe in the case of 16-QAM (see BER plots in Fig. \ref{fig:16_qam_fixed} for 100 predictions). This constrains the number of predictions to be conservatively fixed at 10 in order to achieve good BER performance, which leads to poor bandwidth efficiency. Motivated by this need and opportunity for improvement, in the following subsection (Sec. \ref{sec3d}), we propose an adaptive scheme that allows the receiver to dynamically adjust the number of future predictions employed in the prediction algorithm in accordance with the operating SNR and Doppler. 
\vspace{-1mm}
\subsection{Adaptive channel prediction}
\label{sec3d}
In the previous subsection, the number of future predictions employed in the prediction algorithm is fixed. Here, we propose to adapt the number of predictions in accordance with the operating SNR and Doppler with a motivation to improve bandwidth efficiency and performance. The idea is to create and use a lookup table consisting of the achieved MSE between the channel predictions and the actual channel coefficients for different number of future predictions, SNRs, and Dopplers. The desired target MSE for a given operating SNR is set to be the MSE between the LMMSE channel estimate in (4) and the actual channel coefficients. For a given operating SNR, Doppler, and target MSE, the number of future predictions to be employed in the predictor algorithm is obtained from the lookup table. This makes the prediction algorithm to adaptively employ different number of future predictions for different operating conditions.

\begin{figure}
\centering
\includegraphics[width=\linewidth]{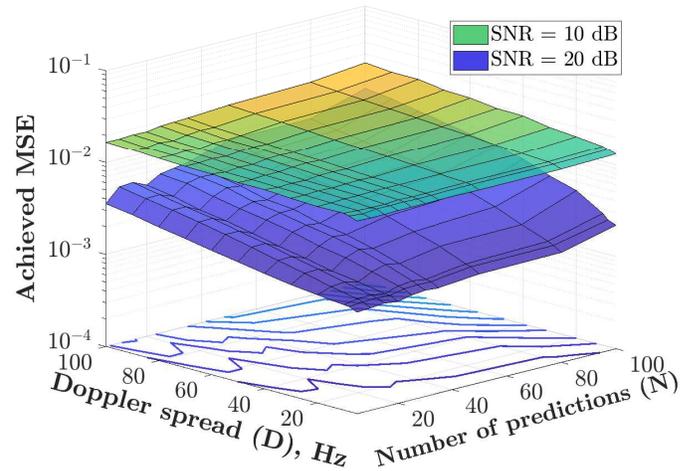}
\caption{Achieved MSE performance of predictions as a function of number of future predictions and Doppler for a given SNR of 10 dB and 20 dB.}
\label{fig:mse_predictor_surf}
\vspace{-4mm}
\end{figure}

In Fig. \ref{fig:mse_predictor_surf}, we show the 3D plots of the entries of the lookup table (i.e., achieved MSE values) for different values of number of predictions ($N=5$ to 100) and Doppler ($D=5$ to 100 Hz) at SNRs of 10 dB and 20 dB. It can be seen that for a fixed $N$, the achieved MSE decreases with decreasing $D$. Likewise, for a fixed $D$, the MSE decreases with decreasing $N$. 
It can also be observed that, for all $N$ and $D$ values, the achieved MSE values at 20 dB SNR are less than those at 10 dB SNR.
The contour lines plotted in the $N$-$D$ plane at the bottom are for the surface corresponding to 20 dB SNR. A given contour line shows all the ($N,D$) values for which the achieved MSE is the same. For example, the outermost contour with $(100,5)$ and $(5,100)$ as the end points has an MSE of 0.002. In this contour, as $D$ decreases from 100 Hz to 5 Hz, $N$ increases from 5 to 100. Further, the innermost contour is a point at $(N,D)=(100,100)$ that has an MSE value of 0.012.

\begin{figure}
    \centering
    \includegraphics[width=\linewidth]{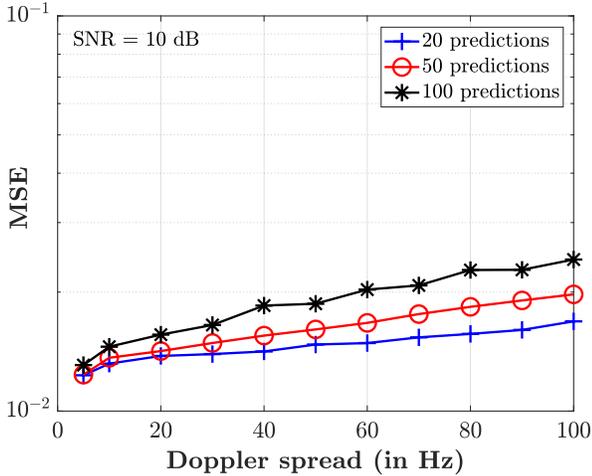}
    \caption{MSE performance of predictions as a function of $f_D$ for a given SNR of 10 dB.}
    \label{fig:mse_vs_dopp}
    \vspace{-4mm}
\end{figure}

In Fig. \ref{fig:mse_vs_dopp}, we show the 2D plots of the MSE performance of the predictions as a function of channel correlation (Doppler spread) for $N=20, 50$, and 100 predictions, at a fixed SNR of 10 dB. 
It is seen that for all values of the Doppler spread, the MSE values are least when the number of predictions is 20 and the highest when the number of predictions is 100. Further, the MSE values across different Doppler spreads are close when $N=100$, while they are even closer when $N=50, 20$. For instance, at 100 Hz Doppler the maximum increase in MSE compared to 10 Hz Doppler is only about 0.011 for the case of 100 predictions, while this number drops to 0.0045 for the case of 20 predictions. This demonstrates that the network trained at 10 Hz is able to generalize and perform quite well across the considered Doppler range (up to 100 Hz).

\begin{figure}[t]
\centering
\includegraphics[width=\linewidth]{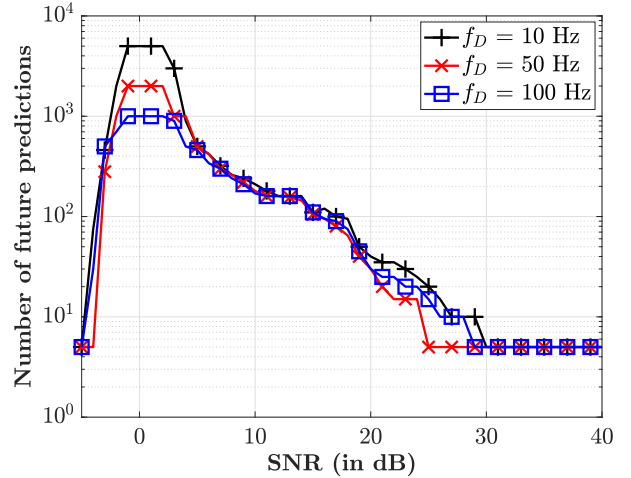}
\caption{Number of future predictions chosen by the prediction algorithm as a function of SNR for different values of $f_D$.}
\label{fig:nfp_vs_snr}
\vspace{-2mm}
\end{figure}

The prediction algorithm chooses the number of predictions (for a given operating SNR, target MSE for that SNR, and Doppler) corresponding to an achieved MSE in the lookup table that is less than the target MSE. Figure \ref{fig:nfp_vs_snr} shows the number of predictions chosen by the algorithm from the lookup table for different SNRs in the range -5 dB to 40 dB and $f_D$ values in the range 10 Hz to 100 Hz. Here, the target MSE at an SNR is obtained by evaluating the MSE of LMMSE estimates obtained by pilot transmissions at that SNR. It can be seen that, for a given $f_D$, the number of predictions chosen by the algorithm shows a bell-shaped behavior as the SNR is increased. For example, at very low SNRs, the algorithm chooses very few number of predictions to meet the target MSE. This is because the correlation in the input to the channel predictor is perturbed significantly by the additive noise having a high variance. This leads to a high MSE of predictions, forcing the algorithm to choose a correspondingly small value for the number of predictions. As the SNR increases, the number of predictions chosen by the algorithm increases. This is because the fluctuations in the correlation in the input decreases with increasing SNR (i.e., decreasing noise variance). As the SNR increases further beyond a certain value, although the disturbance to the correlation reduces further, the algorithm chooses smaller and smaller number of predictions, which can be explained as follows. First, the target MSE (obtained from \eqref{lmmse_est}) decreases with increasing SNR. Second, the achieved MSE for a fixed number of predictions, i.e., the prediction error does not decrease as fast as the target MSE with SNR. The combined effect of these two makes the algorithm to choose reduced number of predictions at high SNR.

\begin{figure*}
\centering
\begin{subfigure}{0.5\linewidth}
\centering
\includegraphics[width=\linewidth]{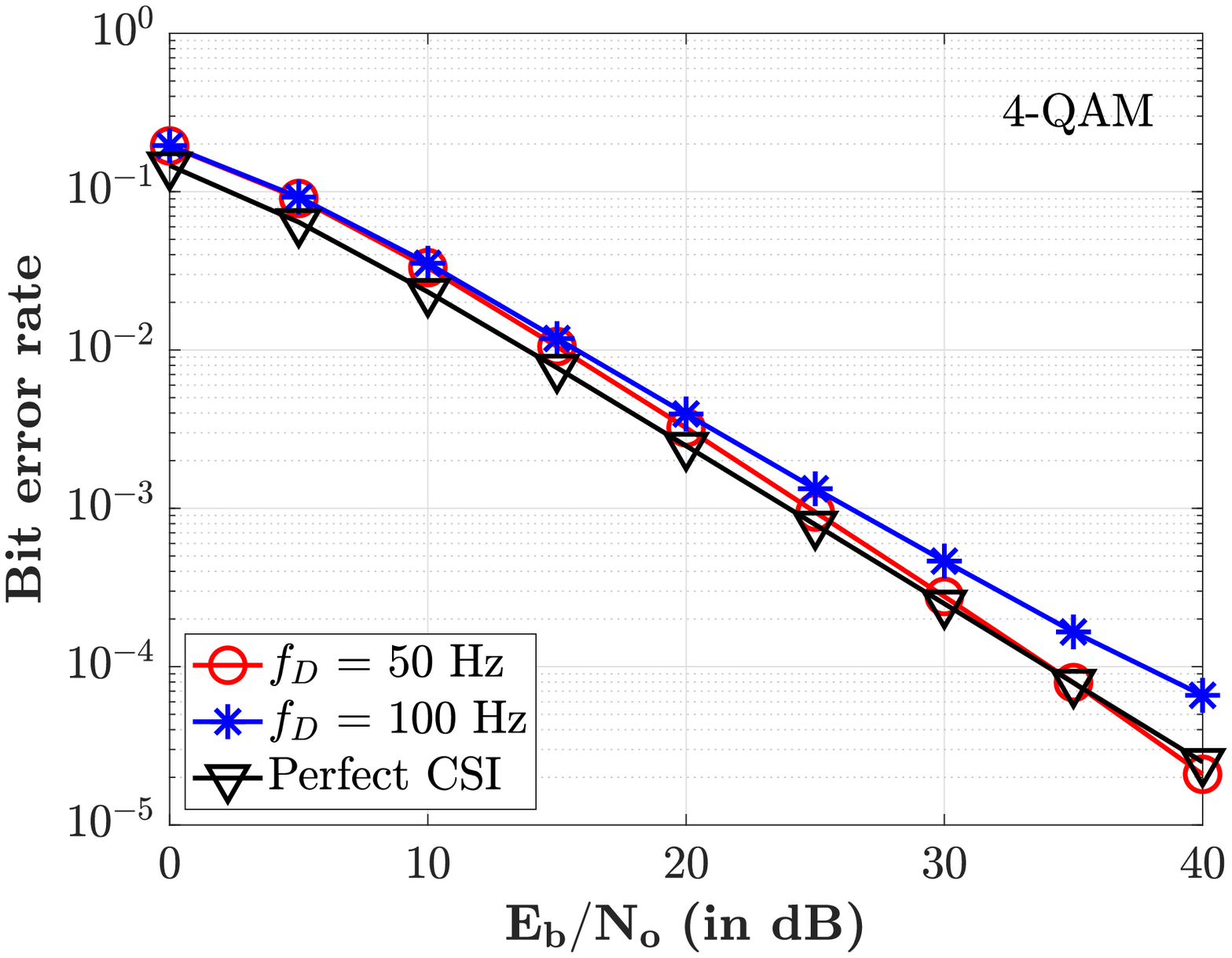}
\caption{4-QAM}
\label{fig:4_qam_nfp}
\end{subfigure}%
\begin{subfigure}{0.5\linewidth}
\centering
\includegraphics[width=\linewidth]{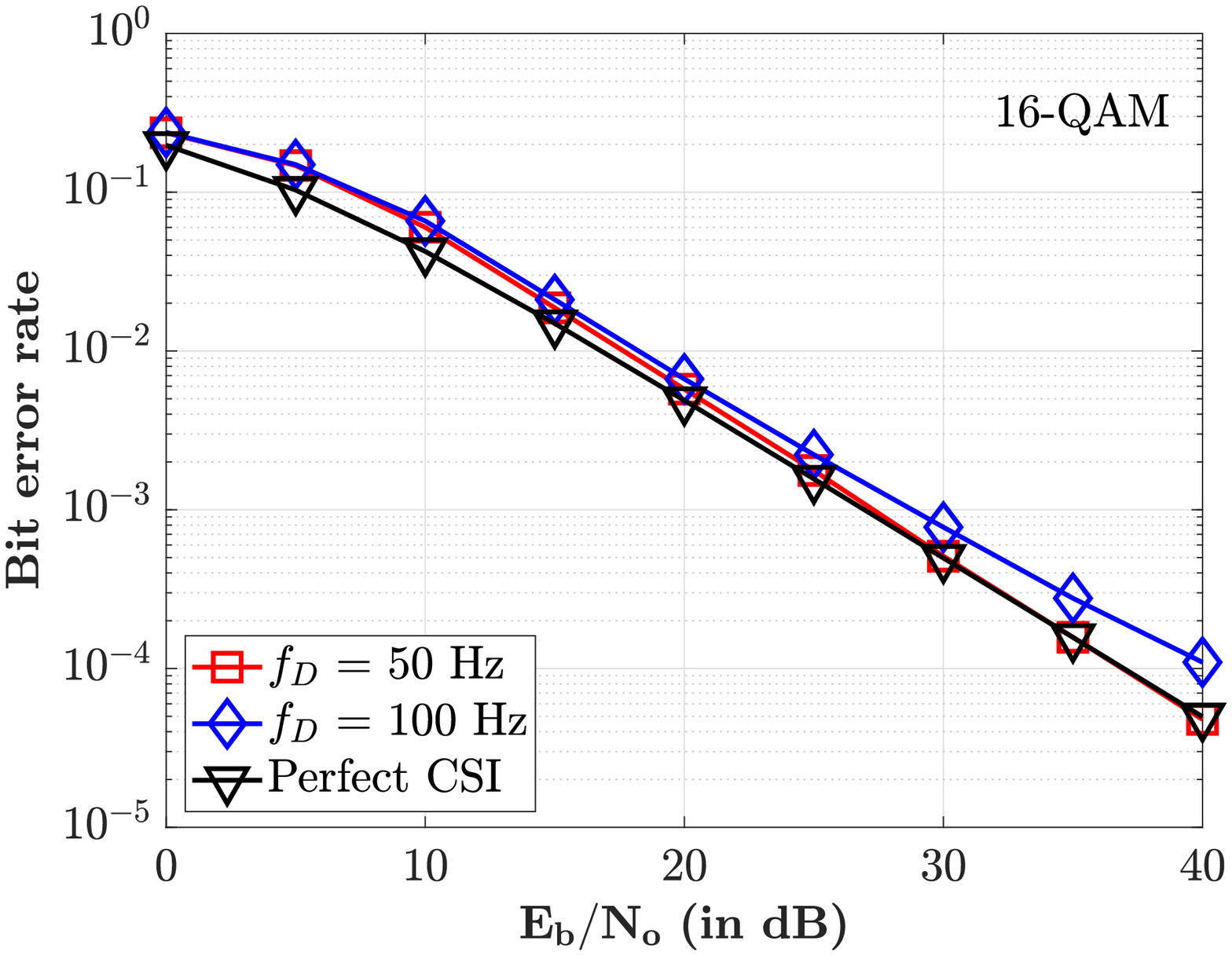}
\caption{16-QAM}
\label{fig:16_qam_nfp}
\end{subfigure}%
\caption{BER performance of the proposed adaptive channel predictor aided receiver with ML decoder at $f_D=50, 100$ Hz for 4-QAM and 16-QAM.}
\label{fig:nfp}
\vspace{-2mm}
\end{figure*}

In Figs. \ref{fig:4_qam_nfp} and \ref{fig:16_qam_nfp}, we demonstrate the BER performance achieved by the proposed adaptive channel predictor aided receiver with ML decoder for 4-QAM and 16-QAM, respectively. The performance at $f_D=50$ Hz and 100 Hz are shown. The performance with perfect CSI is also shown. It can be seen that, with the proposed adaptation of the number of predictions, the receiver is able to achieve a performance that is very close to the ideal performance with perfect CSI for both 50 Hz and 100 Hz Doppler, while also being bandwidth efficient. For example, in Fig. \ref{fig:16_qam_fixed}, for $f_D = 100$ Hz, the greedy scenario fixes the number of predictions to be 100 for all values of $E_b/N_0$, which causes the BER to floor at $3 \times 10^{-2}$. The adaptive scheme, on the other hand, chooses the number of predictions to be 100 until an $E_b/N_0$ value of 7 dB, after which it reduces the number of predictions towards 5 at 40 dB $E_b/N_0$, leading to better performance (no flooring is seen). Likewise, in Fig. \ref{fig:4_qam_fixed}, for $f_D = 50$ Hz, the conservative scenario fixes the number of predictions to be 10 throughout the $E_b/N_0$ range. Although the performance for $f_D=50$ Hz in Figs. \ref{fig:4_qam_fixed} and Fig. \ref{fig:4_qam_nfp} are almost same, the bandwidth efficiency in Fig. \ref{fig:4_qam_fixed} is only 50\%, as there are 10 predictions made for 10 pilots transmitted. On the other hand, in Fig.\ref{fig:4_qam_nfp},  the adaptive scheme chooses the number of predictions to be greater than 10 until the $E_b/N_0$ value of 20 dB. For example, at 20 dB, the number of predictions chosen is 15 which translates to a bandwidth efficiency of 60\%, and at 6 dB, the number of predictions is 100, which achieves a bandwidth efficiency of 90.9\%. 

Next, we consider a non-neural network based benchmarking scheme to compare the performance of the proposed adaptive scheme. The benchmarking scheme employs LMMSE channel estimation and linear interpolation (LI) along with ML decoding. For fair comparison, the bandwidth efficiency is kept same in both the proposed as well as the benchmarking schemes. We achieve this as follows. In both the schemes, transmission is made in frames consisting of pilot symbols and data symbols. The number of pilot symbols ($n_p$) and data symbols ($n_d$) in each frame are taken to be $n_p=10$ and $n_d=N_c$, where $N_c$ is the number of predictions chosen by the predictor algorithm. In the proposed scheme, $n_p=10$ pilot symbols followed by $n_d=N_c$ data symbols are transmitted in a frame. In the benchmarking scheme, one pilot symbol is sent followed by $\frac{n_d}{n_p}$ ($=\frac{N_c}{10}$) data symbols and this pilot-data symbol sequence is repeated till the end of the frame. LMMSE channel estimation is performed during the pilot symbols and linear interpolation is performed to obtain the channel estimates for the duration between two pilot symbols.

\begin{figure}
\centering
\includegraphics[width=\linewidth]{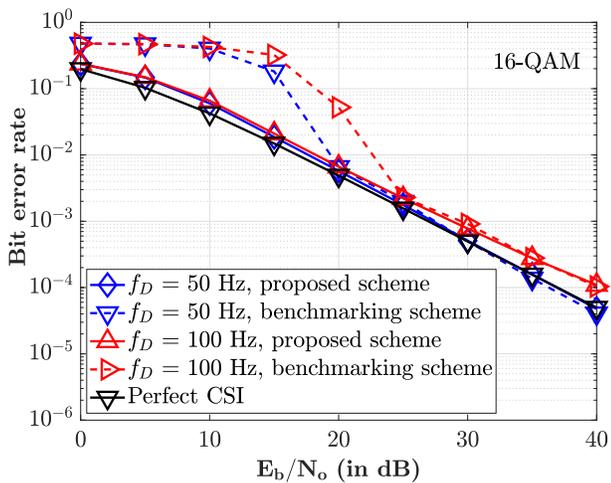}
\caption{BER performance comparison between the proposed adaptive scheme with ML decoder and the benchmarking scheme with LMMSE channel estimation and linear interpolation.}
\label{fig:lmmse_interp}
\vspace{-7mm}
\end{figure}

Figure \ref{fig:lmmse_interp} shows a BER performance comparison between the proposed adaptive scheme with ML decoder and the benchmarking scheme for 16-QAM at $f_D=50$ Hz and 100 Hz. It is seen that the proposed scheme performs significantly better than the benchmarking scheme in the low-to-moderate range of $E_b/N_0$ values (0 to 20 dB). This is because of poor interpolation accuracy in the benchmarking scheme in this $E_b/N_0$ range, which can be explained as follows. The $N_c$ values chosen in the 0 to 20 dB range are large compared to the number of pilot symbols $n_p$ (e.g., $N_c$ is 60 at $E_b/N_0=10$ dB for $f_D = 50$ Hz and $n_p$ is 10). A large value of $\frac{N_c}{n_p}$ means the pilots in a frame are spaced far apart leading to less accurate interpolation. In the higher range of $E_b/N_0$ values, the $\frac{N_c}{n_p}$ ratio becomes small due to smaller values of $N_c$, leading to closer spacing of pilots and hence better interpolation accuracy. This makes the benchmarking scheme perform close to the performance of the proposed scheme in the high $E_b/N_0$ range.

\begin{figure}
    \centering
    \includegraphics[width=\linewidth]{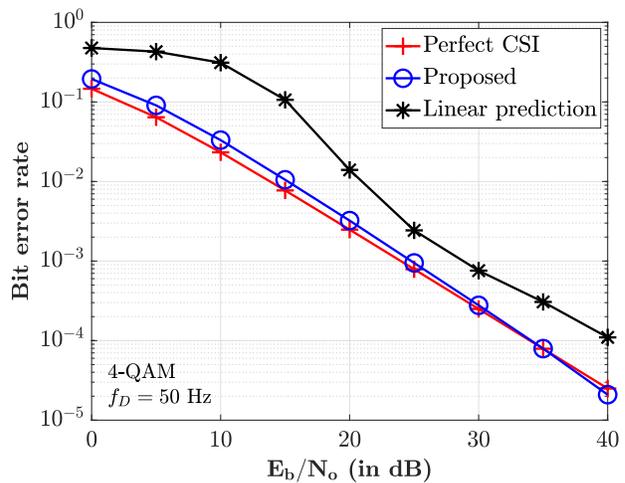}
    \caption{BER performance comparison between the proposed channel predictor aided receiver and the linear prediction aided receiver, both with ML decoder.}
    \label{fig:linear_pred}
    \vspace{-2mm}
\end{figure}

\subsection{Comparison with linear prediction scheme}
\label{sec3e}
In this subsection, we compare the performance of the proposed channel predictor aided receiver with that of a receiver with channel predictor replaced by a linear prediction algorithm. A time-varying channel with $f_D=50$ Hz is considered. The linear prediction algorithm models the time-varying channel coefficients as an auto-regressive (AR) process of order 2, i.e., for any time $t$,
\begin{align}
h(t) = \rho_1 h(t-1) + \rho_2 h(t-2),
\label{ar_process}
\end{align}
where $\rho_1$ and $\rho_2$ are the parameters of the AR(2) process that need to be estimated. The values of $\rho_1$ and $\rho_2$ are computed as follows. 10 pilot symbols followed by $N_c$ 4-QAM data symbols (corresponding to $\eta=\frac{N_c}{10+N_c}$) are transmitted, where $N_c$ (the number of predictions) is chosen in accordance with the adaptive prediction algorithm in Sec. \ref{sec3d}. 10 LMMSE channel estimates at $t=0,1,\cdots,9$ are obtained using the received pilot symbols. A set of 8 equations corresponding to time $t=2,3,\cdots,9$ is obtained from \eqref{ar_process} using the LMMSE estimates. The Yule-Walker (YW) estimation technique \cite{yule},\cite{walker} is employed on these equations to determine the values of $\rho_1$ and $\rho_2$. To obtain $N_c$ channel predictions for $t>9$, \eqref{ar_process} is recursively used with the estimated $\rho_1$ and $\rho_2$ values.

Figure \ref{fig:linear_pred} shows the BER performance comparison between the proposed channel predictor aided receiver and linear prediction aided receiver, both with ML decoder. 
At low SNRs, the LMMSE estimates are noisy and the values of $\rho_1$ and $\rho_2$ obtained through YW estimation theory are inaccurate. This leads to poor quality of predictions and poor BER performance. As SNR increases, the performance of linear prediction algorithm aided receiver improves owing to better $\rho_1$ and $\rho_2$ estimates and reduced $N_c$. However, it is observed that the performance of the proposed channel predictor aided receiver is better than the linear predictor counterpart (e.g., at $10^{-4}$ BER, the proposed predictor aided receiver has an advantage of about 2.5 dB compared to the linear prediction receiver).

\begin{table}
\vspace{6mm}
    \centering
    \begin{tabular}{|p{0.6cm}|p{3.2cm}|p{3.2cm}|}
        \hline
         \textbf{Model} & \textbf{Tap delays (ns)} & \textbf{PDP (in dB)} \\
         \hline
         EPA & 0, 30, 70, 90, 110, 190, 410 & 0, -1, -2, -3, -8, -17.2, -20.8\\
         \hline
         EVA & 0, 30, 150, 310, 370, 710, 1090, 1730, 2510 & 0, -1.5, -1.4, -3.6, -0.6, -9.1, -7, -12, -16.9\\
         \hline
         ETU & 0, 50, 120, 200, 230, 500, 1600, 2300, 5000 & -1, -1, -1, 0, 0, 0, -3, -5, -7\\
         \hline
    \end{tabular}
    \caption{Tap profile of the 3GPP channel models.}
    \label{tab:3gpp_models}
\end{table}

\begin{figure}
    \begin{subfigure}{0.49\linewidth}
        \centering
        \includegraphics[width=4.75cm, height=4.5cm]{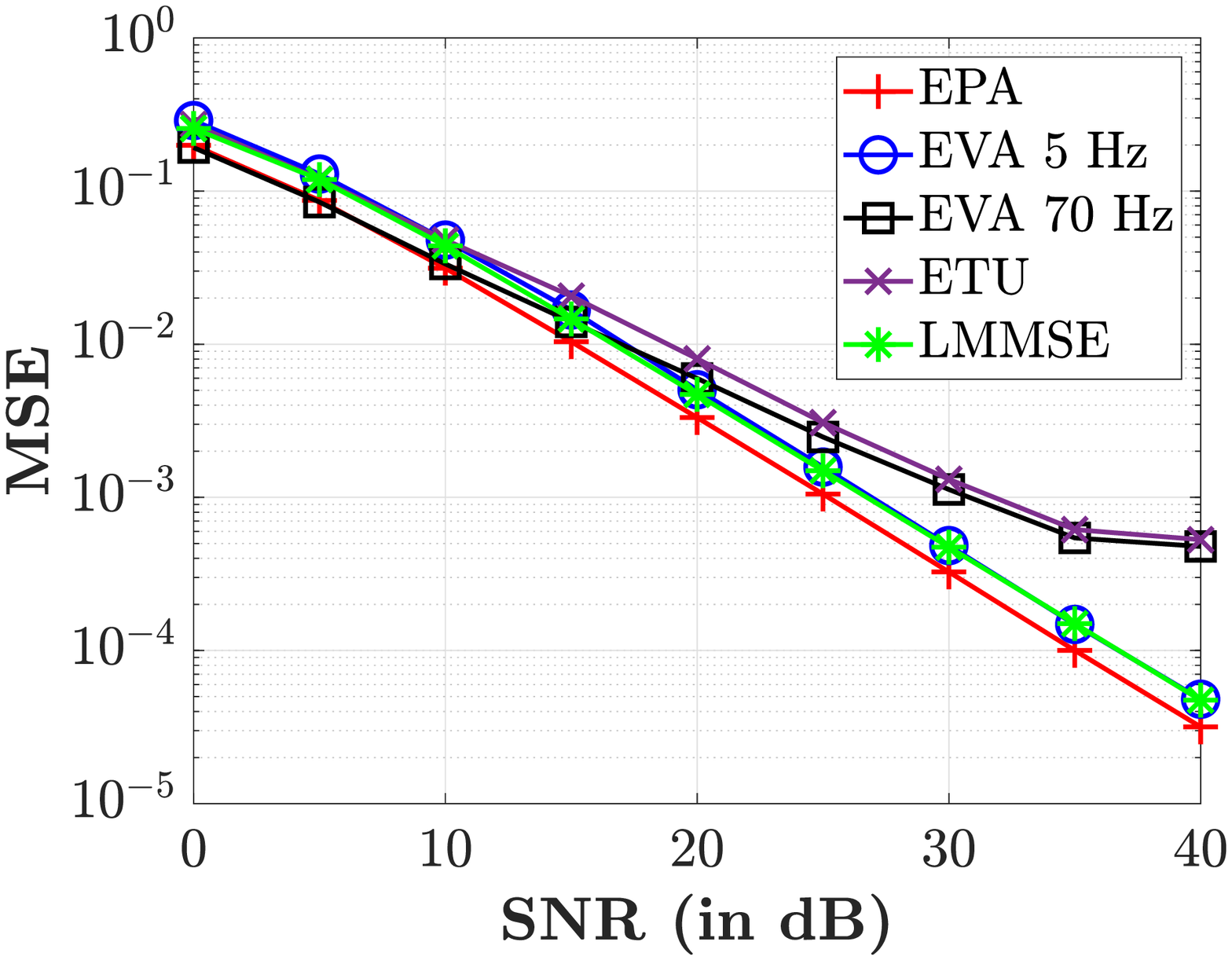}
        \caption{SNR}
        \label{fig:3gpp_ber_vs_snr}
    \end{subfigure}
    \begin{subfigure}{0.49\linewidth}
        \centering
        \includegraphics[width=4.75cm, height=4.5cm]{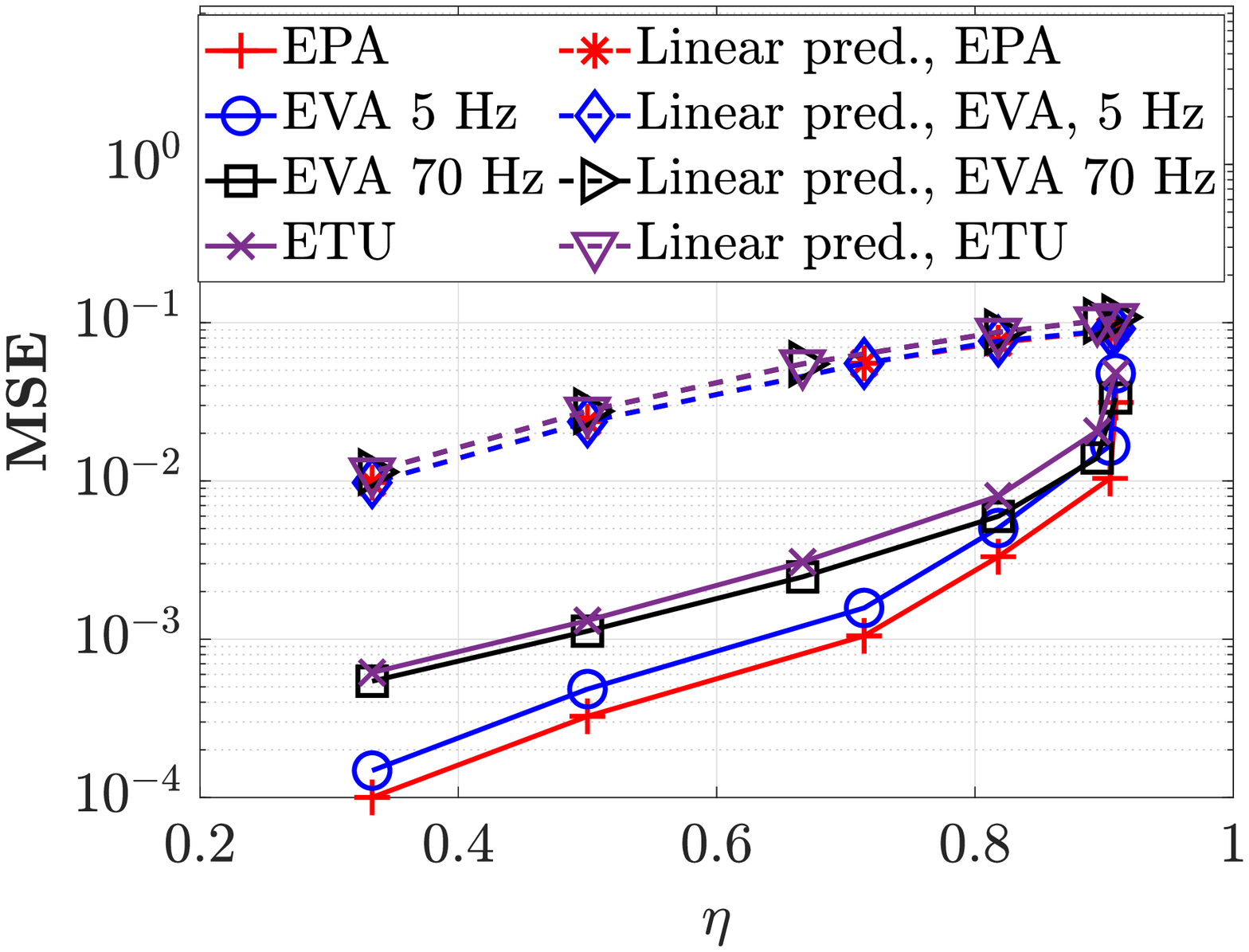}
        \caption{$\eta$}
        \label{fig:3gpp_mse_vs_eff}
    \end{subfigure}
    \caption{MSE performance of the proposed channel predictor network under various 3GPP channel models in \cite{3gpp_1},\cite{3gpp_2} as a function of SNR and $\eta$.}
    \label{fig:3gpp}
\end{figure}

\subsection{Performance in 3GPP channel models}
\label{sec3f}
In this subsection, we present the MSE performance of the proposed channel predictor network under different multipath channel propagation models defined by 3GPP \cite{3gpp_1},\cite{3gpp_2}.
We consider extended pedestrian A (EPA) model, extended vehicular A (EVA) model, and extended typical urban (ETU) model under slow ($f_D=5$ Hz) and fast ($f_D=70$ Hz) mobility conditions. The tap delays and power delay profiles of these models are given in Table \ref{tab:3gpp_models}. We use $f_D=5$ Hz for EPA model, $f_D = 5,70$ Hz for EVA model, and $f_D = 70$ Hz for ETU model. The MSE performance in a multipath propagation model with $L$ taps is obtained as follows. For estimating the channel across multiple taps, $n_p$ pilot sequences are transmitted. Each pilot sequence consists of a pilot symbol along with $L-1$ preceding and succeeding zeroes, making the length of each pilot sequence to be $2L-1$. $n_p$ LMMSE channel estimates corresponding to each tap are obtained from the received pilot sequences. $N_c$ number of deep channel predictions are made on each tap and the MSE of the predicted coefficients are calculated with respect to the actual channel coefficients.

Training of the predictor network is carried out using channel coefficients obtained from the synthetic dataset obtained for a single-tap channel (i.e., the network is not trained with the dataset from the actual 3GPP models). However, the trained network could work well for all the actual 3GPP models which have multiple taps and non-uniform power-delay profiles as shown in Table \ref{tab:3gpp_models}. 
This can be seen in Fig. \ref{fig:3gpp_ber_vs_snr} which shows the obtained MSE values as a function of SNR with $n_p=10$ and $N_c$ is chosen according to algorithm in Sec. \ref{sec3d}. It is seen that the MSE values for all the considered 3GPP models decrease with SNR, closely following the MSE of the LMMSE estimates in the low and mid SNR regimes. For EVA with $f_D=70$ Hz and ETU, there is a small deviation observed in MSE in the high SNR regime due to high Doppler spread. 
Figure \ref{fig:3gpp_mse_vs_eff} shows the MSE performance of the channel predictor as a function of the bandwidth efficiency $\eta$ for the 3GPP models. It is seen that the MSE is below $10^{-2}$ for $\eta \leq 0.8$, showing that the predictions are reasonably accurate even when operating at a bandwidth efficiency of 80\%. On the other hand, the MSE achieved by the linear prediction scheme in Sec. \ref{sec3e} is found to be much higher.
So, although the predictor network is trained on a synthetic dataset, the network could learn to observe the correlation in the channel coefficients at its input and use the learnt correlation to make further predictions, even in settings or environment not seen while training.
This demonstrates the generalization capabilities and robustness of the proposed channel predictor.

\begin{figure}
    \centering
    \includegraphics[width=\linewidth]{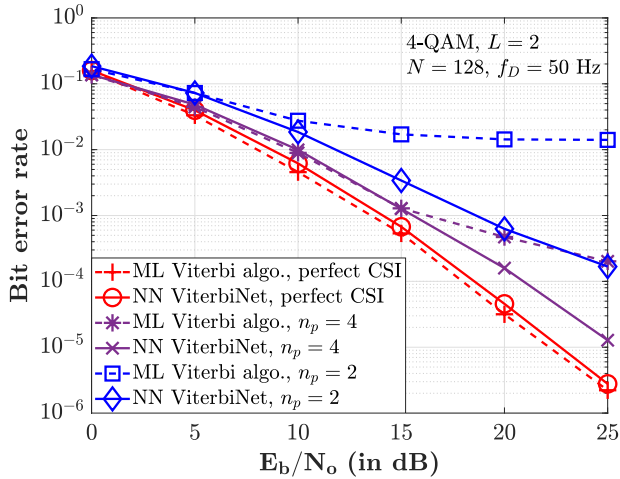}
    \caption{BER performance of the proposed predictor network in a CPSC system with NN-based ViterbiNet detector.}
    \label{fig:cpsc}
    \vspace{-4mm}
\end{figure}

\subsection{Block transmission in doubly-selective fading channel}
\label{sec3g}
In this subsection, we consider block transmission and detection in doubly-selective fading channels and evaluate the performance of the proposed deep channel predictor. We consider a cyclic prefix single carrier (CPSC) system, where the channel is taken to be both frequency and time selective. Each CPSC frame consists of $n_p$ pilot sequences (see Sec. \ref{sec3f}) followed by $N_c = N-n_p(2L-1)$ data symbols, where $N$ is the CPSC frame length, and $L$ is the number of channel taps. Deep channel prediction is done on each tap and the predicted coefficients are given, along with the received data symbols, as input to the detector. We demonstrate the advantage of using NN-based detection in such channels by comparing the performance of a) maximum-likelihood (ML) detection using Viterbi algorithm and b) NN-based detection.  In the ML detection using Viterbi algorithm, the channel coefficients predicted by the deep channel predictor are used to evaluate the likelihood costs. For NN-based detection, we use ViterbiNet \cite{shlezinger}, which uses learning based computation of likelihoods in the Viterbi algorithm. We train the ViterbiNet detector using the fade coefficients predicted by the deep channel predictor.

Figure \ref{fig:cpsc} shows the BER performance of the considered CPSC system with $N=128$, $L=2$,  4-QAM, and $f_D=50$ Hz. The performance of ML Viterbi detector and NN-based ViterbiNet detector are shown. Performance with channel prediction for $n_p=2, 4$ per CPSC frame are shown. Performance plots with perfect CSI are also shown for comparison. The following observations can be made from Fig. \ref{fig:cpsc}. The ViterbiNet detector trained using perfect CSI achieves almost the same performance as the ML Viterbi detector performance with perfect CSI. The performance of both the detectors degrade when predicted channel coefficients are used. The performance degradation in ML Viterbi detector is significantly higher than that in ViterbiNet detector. For example, the ML Viterbi detector performance floors at a BER of about $10^{-2}$ for $n_p=2$ at 25 dB SNR, whereas the ViterbiNet detector achieves a significantly better BER of about $10^{-4}$ for the same SNR. Also, for $n_p=4$, the ViterbiNet detector performs close to that with perfect CSI (within about 2.5 dB gap at $10^{-5}$ BER), whereas ML Viterbi detector starts flooring at $10^{-4}$ BER itself. This is in corroboration with the results reported in \cite{shlezinger}, where it is shown that, in the presence of imperfect channel state information (CSI), the performance of conventional Viterbi algorithm degrades significantly whereas the NN-based ViterbiNet detection achieves significantly better performance. The better performance of the combination of the proposed deep channel prediction and NN-based ViterbiNet detection therefore demonstrates the benefit of learning approach in communication receivers.

\begin{figure*}[ht]
\centering
\includegraphics[width=0.8\linewidth]{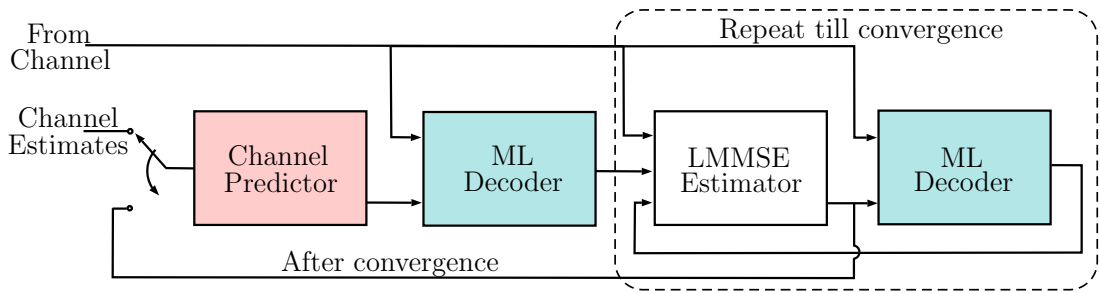}
\caption{Block diagram of the proposed data driven channel prediction scheme.} 
\label{fig:predictor_block_with_feedback}
\vspace{-0mm}
\end{figure*}

\begin{figure}
\centering
\includegraphics[width=0.8\linewidth]{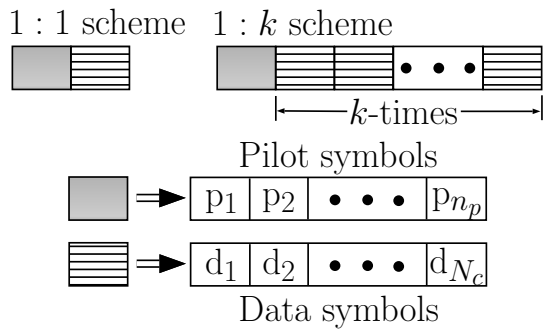}
\caption{Arrangement of pilot and data symbols in data driven channel prediction.}
\label{fig:timing_dia}
\vspace{-2mm}
\end{figure}

\section{Data driven channel prediction}
\label{sec4}
In this section, we present the proposed data decision driven channel prediction architecture and its performance. The motivation for the data decision driven approach is as follows. We note that the maximum bandwidth efficiency obtained in the adaptive prediction scheme proposed in the previous section is 90.9\%, which is obtained when the number of predictions $N_c=100$ and number of pilots $n_p=10$. In the high SNR region, however, the algorithm reduces $N_c$ to 5, where it attains a bandwidth efficiency of only 33\%. We aim to improve this low bandwidth efficiency by using a data decision driven prediction architecture proposed in the following subsection.

\begin{figure*}
\centering
\begin{subfigure}{0.5\linewidth}
\centering
\includegraphics[width=\linewidth]{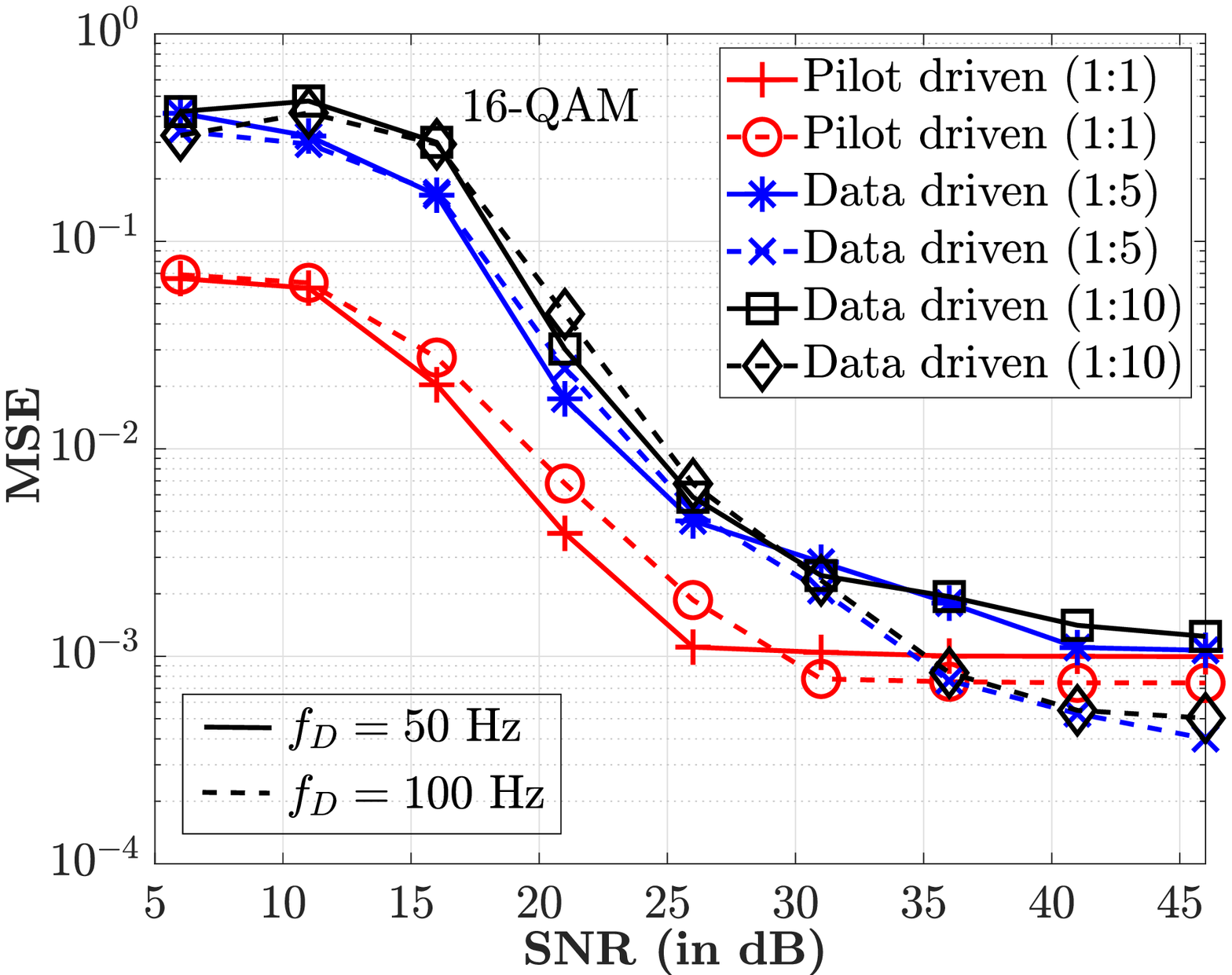}
\caption{MSE}
\label{fig:16_qam_mse_feedback}
\end{subfigure}%
\begin{subfigure}{0.5\linewidth}
\centering
\includegraphics[width=\linewidth]{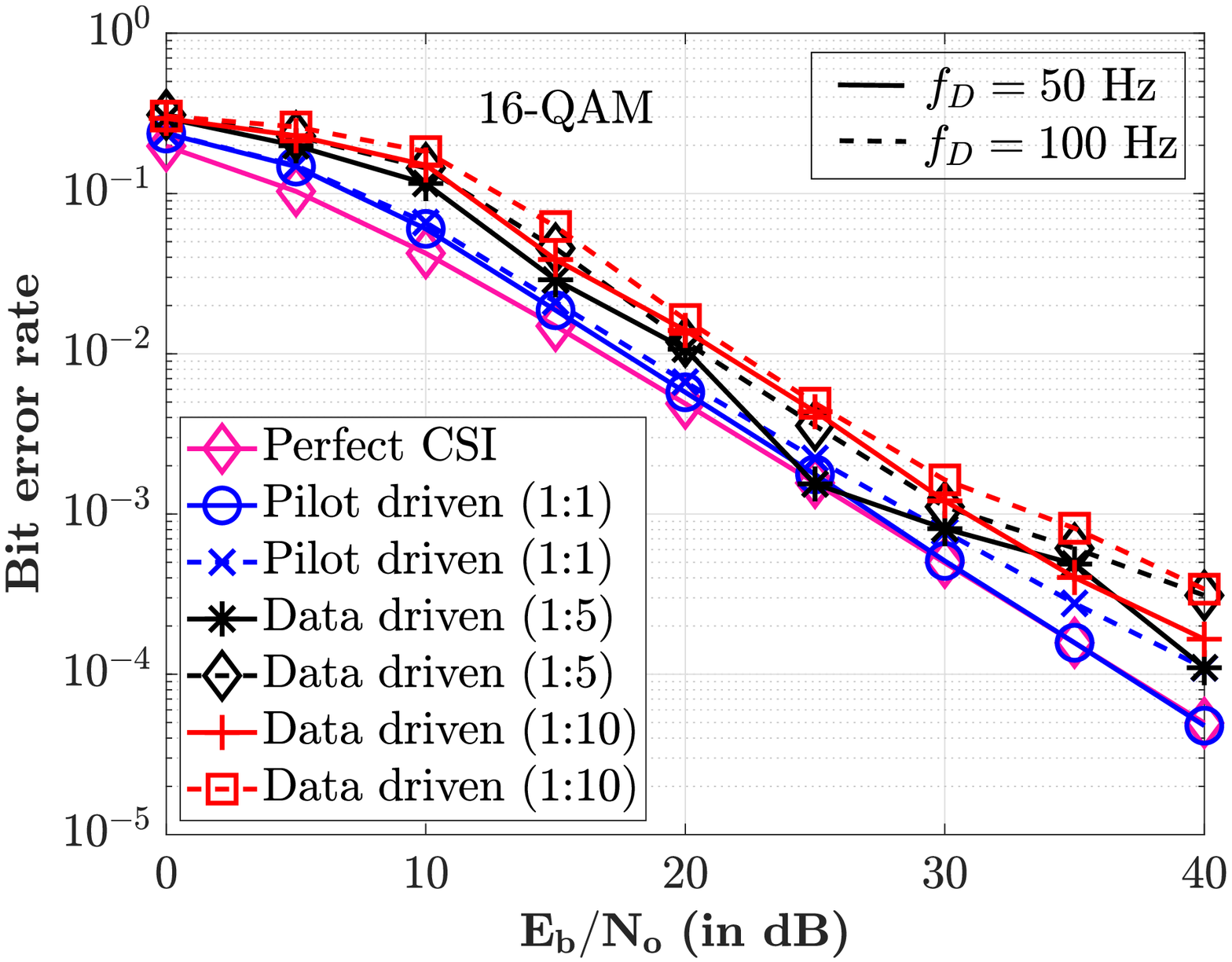}
\caption{BER}
\label{fig:16_qam_feedback}
\end{subfigure}%
\caption{MSE and BER performance of 1:$k$ data decision driven channel prediction scheme with ML decoder at $f_D = 50, 100$ Hz for 16-QAM.}
\label{fig:feedback_mse_perf}
\vspace{-3mm}
\end{figure*}

\subsection{Architecture} 
\label{sec4a}
In the proposed data driven prediction approach, we adopt a 1:$k$ transmission scheme in which 1 pilot block (consisting of $n_p$ pilot symbols) is sent every $k$ data blocks (each data block consisting of $N_c$ data symbols) as shown in Fig. \ref{fig:timing_dia}. The shaded block in Fig. \ref{fig:timing_dia} represents a pilot block which is used to obtain LMMSE estimates of the channel. The predictions obtained using these estimates are used to decode $N_c$ data symbols transmitted in the subsequent striped data block. In the case of $k=1$ (i.e., 1:1 scheme), one pilot block and one data block are sent in an alternating fashion, leading to a bandwidth efficiency of $\frac{N_c}{n_p+N_c}$. The 1:1 scheme is a purely pilot driven prediction scheme and there is no data driven prediction. On the other hand, for $k>1$, there is data decision driven prediction (described in the next paragraph) and the bandwidth efficiency improves to $\frac{kN_c}{n_p+kN_c}$. 

The block diagram of the proposed data decision driven prediction architecture is shown in Fig. \ref{fig:predictor_block_with_feedback}. The channel predictor and ML decoder blocks are the same as in Fig. \ref{fig:without_feedback_block_dia}. The predictions from the channel predictor are fed to the ML decoder, which uses the predictions to decode data symbols received through the channel. The LMMSE estimator block receives these decoded symbols from the ML decoder along with the symbols received through the channel. Here, the decoded symbols from the ML decoder are treated as pilots and the signal received from the channel as the faded version of these pilots, and an LMMSE estimate of the fade coefficients are obtained from this decoded data. These LMMSE channel estimates act as a refined version of the predictions made by the channel predictor network. The refined channel estimates are used to once again decode the data symbols using a second ML decoder. If the decoded symbols from the second ML decoder match the decoded symbols from the first ML decoder, then the refined channel estimates are fed back to the input of the channel predictor to enable further predictions. If they do not match, then the current decoded symbols are fed back to the LMMSE estimator (as pilots) as before, and another set of refined channel estimates are obtained. This process of data decoding and channel estimate refinement are iteratively repeated until the decoded outputs from two consecutive iterations match. When this happens, the LMMSE channel estimates in the subsequent iterations do not change, as the decoded symbols being fed back as pilots to the LMMSE estimator do not change. This is set as the convergence criterion in the receiver. If the criterion is not met for a certain number of iterations (e.g., 200 iterations), then the last obtained LMMSE estimates are used as the feedback to the input to make further predictions. If there is an error in decoding a symbol and the corresponding LMMSE estimate is fed back to the channel predictor input, then the further predictions obtained may have a large MSE and this may result in more subsequent errors. The value of $k$ is chosen such that this error propagation is minimized.

\subsection{Performance results}
\label{sec4b}
In Fig. \ref{fig:16_qam_mse_feedback}, we present the MSE performance of the predictions made by the data driven channel prediction scheme at $f_D=50,100$ Hz for 16-QAM. The values of $k$ considered are $k=1,5,10$. It is seen that at low SNR values, the 1:1 scheme has a lower MSE than the 1:$k$ schemes, $k=5,10$. This is because there is no data driven prediction in the 1:1 scheme and hence there is no error propagation due to decoding errors. On the other hand, in the 1:$k$ scheme ($k=5,10$), the MSE of the predictions degrades due to error propagation caused by decoding errors at these low SNR values. As the SNR increases, the MSE of the 1:$k$ schemes decreases (due to fewer decoding errors) and the gap from the MSE of 1:1 scheme reduces. In the high SNR regime, the 1:1 and 1:$k$ schemes achieve similar MSE performance, again due to fewer decoding errors in the 1:$k$ schemes.

Figure \ref{fig:16_qam_feedback} shows the achieved BER performance with ML decoder corresponding to the MSE performance presented in Fig. \ref{fig:16_qam_mse_feedback}. We observe that both the 1:5 and 1:10 schemes perform close to the 1:1 scheme. We also see that the BER performance of 1:5 scheme is closer to that of the 1:1 scheme than the 1:10 scheme for both $f_D$ values, which is justified owing to larger number of pilots in the 1:5 scheme. We further note that the main advantage of the data driven prediction scheme is its bandwidth efficiency due to the reduced number of pilots used in the scheme. For example, for the 1:10 scheme, when the maximum value of $N_c = 100$ is chosen by the adaptive algorithm, the total number of symbols decoded per estimation phase (consisting of 10 pilot transmissions) is $10^3$ (i.e., 10 prediction phases with 100 symbols per prediction phase) and the bandwidth efficiency achieved is $\frac{1000}{1010}=99$\%. Likewise, for $N_c = 5$, the bandwidth efficiency achieved is $\frac{50}{60}=83.3$\%. Similarly, for the 1:5 scheme, the maximum and minimum achieved bandwidth efficiencies are 98\% and 71.4\%, respectively. Recall that, in the previous scheme without data driven prediction (i.e., 1:1 scheme), the bandwidth efficiencies achieved for $N_c=100$ and 5 are 90.9\% and 33\%, respectively. In conclusion, we find that the system is able to utilize the channel very efficiently by maximizing the number of data symbol transmission phases per pilot symbol transmission phase, and this is achieved at the cost of a small loss in BER performance.

\subsection{Comparison with NN-based prediction scheme in \cite{madhubabu}}
\label{sec4c}

\begin{table}
\centering
\begin{tabular}{|l|l|l|l|}
\hline
\textbf{Scheme} & $n_p$ & $k$ & \textbf{$N_c$}\\
\hline
10:40 & 10 & 5 & 8 \\
\hline
50:200 & 50 & 5 & 40 \\
\hline
100:400 & 100 & 5 & 80 \\
\hline
150:600 & 150 & 10 & 60 \\
\hline
\end{tabular}
\vspace{2mm}
\caption{Values of $n_p$, $k$, and $N_c$ used for comparison with NN-based prediction scheme in \cite{madhubabu}.}
\label{tab:k_nc_perf_comp}
\vspace{-4mm}
\end{table}

\begin{figure}
\vspace{-2mm}
\centering
\includegraphics[width=\linewidth]{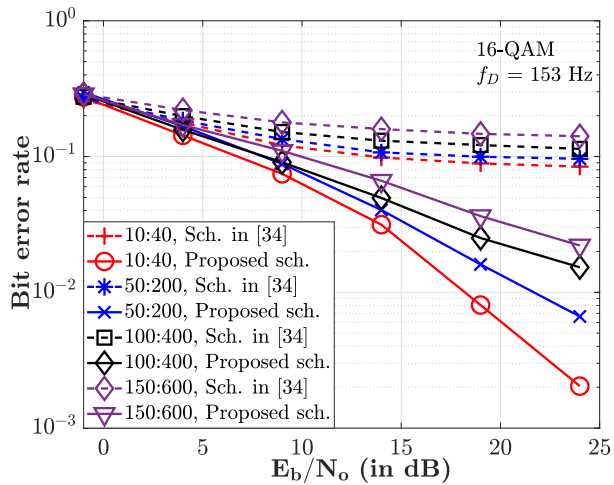}
\caption{BER performance comparison between the proposed scheme and the NN-based prediction scheme in \cite{madhubabu}, both with ML decoder.}
\label{fig:model_comp}
\vspace{-4mm}
\end{figure}

In this subsection, we compare the performance of the proposed data driven channel prediction scheme with an LSTM based channel prediction scheme reported in \cite{madhubabu} both with ML decoder. 16-QAM modulation and $f_D=153$ Hz are considered. We fix the ratio of number of pilot symbols ($n_p$) to the number of data symbols (also the number of predictions, $kN_c$ (see Sec. \ref{sec4a})), while varying $n_p$ and $kN_c$. The ($n_p$:$kN_c$) values considered are (10:40), (50:200), (100:400), and (150:600). Table \ref{tab:k_nc_perf_comp} shows the values chosen for $n_p$, $k$, and $N_c$ in each case. Figure \ref{fig:model_comp} shows the BER comparison between the two schemes. As expected, the performance of (10:40) scheme is better than the (150:600) scheme in both the cases due to smaller number of predictions per pilot block. It is further observed that the proposed scheme achieves significantly better BER performance compared to the scheme in \cite{madhubabu}. This performance advantage in the proposed scheme is attributed to the data driven feature and the training enhancement features incorporated in the proposed scheme.

\section{Conclusion}
\label{sec5}
We proposed a neural network based framework for the design of robust receivers in time-varying fading channels with temporal correlation in the fading process. Central to the proposed framework is the deep channel predictor which uses an RNN that learns the underlying correlation model in the fading process and makes predictions of the channel fade coefficients into the future thereby reducing pilot resources. An FCNN based data symbol decoder aided by the RNN based channel predictor constituted the receiver architecture. The basic version of the channel predictor kept the number of future predictions fixed regardless of the operating SNR and Doppler. An augmented adaptive channel prediction architecture which chose the number of future predictions in accordance with the operating SNR and Doppler further improved the bandwidth efficiency and performance. A data decision driven prediction architecture with decision feedback provided a balance between pilot resources and performance. The achieved robustness in the receiver performance over a range of Doppler and SNR conditions demonstrates that the proposed deep channel prediction approach is a promising approach for receiver design in time-varying fading channels. Finally, we note that the deep channel prediction considered in this paper is in the time-domain. Accordingly, we presented the performance of the proposed predictor in CPSC systems which are essentially  time-domain systems. Learning architectures for channel prediction in frequency-domain systems such as OFDM systems can be devised likewise. We suggest this as a topic for future research. Deriving theoretical guarantees for the performance of deep neural networks based approaches is known to be often intractable and difficult. This could be an important focus area for future research.

\ifCLASSOPTIONcaptionsoff
\newpage
\fi

\end{document}